\documentclass[iop,numberedappendix,twocolappendix]{emulateapj}
\usepackage{natbib}
\usepackage{amsmath}
\usepackage[dvipsnames]{xcolor}
\graphicspath{{img/}}

\newcommand{\Hy}{\ensuremath{\textrm{H}}}
\newcommand{\HI}{\ensuremath{\textrm{H} \, \textsc{i}}}
\newcommand{\HII}{\ensuremath{\textrm{H} \, \textsc{ii}}}

\newcommand{\HeI}{\ensuremath{\textrm{He} \, \textsc{i}}}
\newcommand{\HeII}{\ensuremath{\textrm{He} \, \textsc{ii}}}
\newcommand{\HeIII}{\ensuremath{\textrm{He} \, \textsc{iii}}}
\newcommand{\nH}{n_{\Hy}}

\newcommand{\mean}[1]{\langle #1 \rangle}
\newcommand{\lyalpha}{Ly$\alpha$}
\newcommand{\Deltaopt}{\Delta_{\bigstar}}
\newcommand{\taue}{\tau_{\rm e}}
\newcommand{\nel}{n_{\rm e}}
\newcommand{\xe}{x_{\rm e}}
\newcommand{\lj}{\lambda_{\rm P}}

\shortauthors{O\~norbe et al.}

\begin{document}
\label{firstpage}

\title{Constraining Reionization with the $z \sim 5-6$ Lyman-$\alpha$ Forest Power Spectrum: The Outlook After Planck}

\author{J.~O\~norbe\altaffilmark{1},
J.~F.~Hennawi\altaffilmark{1,2},
Z.~Luki\'c\altaffilmark{3},
M.~Walther\altaffilmark{1,2}}

\altaffiltext{1}{Max-Planck-Institut f\"ur Astronomie, Königstuhl 17, 69117 
  Heidelberg, Germany}
\altaffiltext{2}{Department of Physics, University of California, Santa Barbara, CA 93106-9530, USA}
\altaffiltext{3}{Lawrence Berkeley National Laboratory, CA 94720-8139, USA}

\begin{abstract}
The latest measurements of cosmic microwave background electron-scattering
optical depth
reported by Planck significantly reduces the allowed space of $\HI$
reionization models, pointing toward a later ending and/or less
extended phase transition than previously believed. Reionization
impulsively heats the intergalactic medium (IGM) to 
    $\sim 10^4\,{\rm  K}$, 
    and owing to long cooling and dynamical times in the diffuse
gas that are comparable to the Hubble time, memory of reionization heating is
retained. Therefore, a late-ending reionization has significant
implications for the structure of the $z\sim 5-6$ \lyalpha{} forest.  Using state-of-the-art hydrodynamical simulations
that allow us to vary the timing of reionization and its associated
heat injection, we argue that extant thermal signatures from
reionization can be detected via the \lyalpha{} forest power spectrum
at $5 < z < 6$. This arises because the small-scale cutoff in the
power depends not only on the IGM temperature at these epochs, but
is also particularly sensitive to the pressure-smoothing scale set by
the IGM full thermal history.  Comparing our different reionization
models with existing measurements of the \lyalpha{} forest
flux power spectrum at
$z=5.0-5.4$, we find that models satisfying Planck's $\tau_e$
constraint favor a moderate amount of heat injection consistent with
galaxies driving reionization, but disfavoring quasar driven
scenarios.  
We study the feasibility of measuring the
flux power spectrum at $z\simeq 6$ using mock quasar spectra and conclude that a
sample of $\sim 10$ high-resolution spectra with an attainable signal-to-noise ratio
will allow distinguishing between different reionization scenarios.
\end{abstract}


\section{Introduction}
\label{sec:Introduction}

How and when the first luminous sources reionized diffuse baryons in the 
intergalactic medium (IGM) is one of the most fundamental open questions in 
cosmology. Recently, the Planck collaboration have released new tighter 
constraints on reionization from cosmic microwave background (CMB) observations 
\citep{Planck:2016a,Planck:2016b}. Using for the first time the low-multipole 
$EE$ data from Planck-HFI, the Planck team has significantly 
improved our constraints on the 
cosmic reionization optical depth, $\taue$, finding 
$\taue=0.058 \pm 0.012$ \citep{Planck:2016b}.

The reionization of $\HI$ by the UV background from galaxies and/or quasars 
results in the highly ionized IGM probed at $z \lesssim 6$ by observations of 
the \lyalpha{} forest \citep{McQuinn:2016b}.
Indeed, 
observations of complete Gunn-Peterson absorption in the spectra of many of the 
highest $z\sim6$ quasars, along with the steep rise of both the \lyalpha{} 
optical depth and its sightline-to-sightline scatter with redshift, has led to 
the consensus that we are witnessing the end of reionization only at $z\sim6$ 
\citep{Fan:2006,McGreer:2015,Becker:2015}.
However, \lyalpha{} opacity can only set lower 
limits on the redshift of reionization $z\gtrsim6$, because the overly 
sensitive \lyalpha{} transition saturates for volume-averaged neutral fractions 
$\mean{x_{\HI}}\gtrsim10^{-4}$, which is far too small to pinpoint the redshift of 
reionization.
While new constraints have emerged from the possible presence of a 
\lyalpha{} IGM damping wing in the highest redshift known quasar at $z=7.1$ 
\citep{Mortlock:2011, Simcoe:2012, Greig:2016}, and the decreasing strength of 
\lyalpha{} emission lines in $z\sim 6-7$ galaxies 
\citep{Caruana:2014,Schmidt:2016,Sadoun:2017}, the resulting constraints on 
$\mean{x_{\rm HI}}$ are degenerate with the intrinsic properties of the 
high-$z$ quasars and galaxies that they have been deduced from. We are in 
need of another technique to probe when reionization occurred.

During reionization, ionization fronts propagate supersonically through the IGM, 
impulsively heating gas to $\sim10^{4}$ K \citep{Abel:1999,Davies:2016}. 
The integrated energy balance of heating and inverse-Compton and 
adiabatic cooling then gives rise to a power law temperature-density relation, $T = 
T_{0} (\rho/\bar{\rho})^{\gamma-1}$ 
\citep{MiraldaEscude:1994,Hui:1997,Hui:2003,Meiksin:2009,McQuinn:2009,
McQuinn:2016}. Another important physical ingredient to describe the thermal 
state of the IGM is the gas pressure support that produces an effective 
three-dimensional smoothing of the baryon distribution relative to the dark 
matter at a characteristic scale, $\lj$. In an expanding universe with an 
evolving thermal state at a given epoch, this scale depends on the entire 
thermal history of IGM because fluctuations at earlier times expand or fail to 
collapse depending on the IGM temperature at that epoch 
\citep{Gnedin:1998,Rorai:2013,Kulkarni:2015,Onorbe:2017,Rorai:2017b}.
At redshift 
$z$, the level of pressure smoothing depends not on the prevailing 
pressure/temperature at that epoch, but rather on the temperature of the IGM in 
the past. The IGM pressure-smoothing scale, $\lj$, thus provides an integrated 
record of the thermal history of the IGM, and is sensitive to the timing of and 
heat injection by reionization events.

Measurements of the statistical properties of the \lyalpha{}
forest are 
sensitive to the thermal state of the IGM through the thermal Doppler broadening 
of absorption lines, as well as the pressure-smoothing. The standard approach 
has been to compare measurements of different statistics to cosmological 
hydrodynamical simulations 
\citep{Zaldarriaga:2001,Theuns:2002a,Viel:2009,Lidz:2010,Becker:2011, 
Garzilli:2012,Rorai:2013,Irsic:2014,Rorai:2017,Rorai:2017b} to deduce the thermal 
parameters (e.g., $T_{0}$, $\gamma$ or $\lj$) that best describe the IGM thermal 
state. 
As a larger number of high-resolution spectra of 
quasars have become available at higher redshifts $z\gtrsim4$, the same approach 
has been applied to study  the thermal state of the IGM at these redshifts, 
where the \lyalpha{} forest is more sensitive to the timing and nature of 
hydrogen reionization 
\citep{Theuns:2002a,Hui:2003,Furlanetto:2009,Cen:2009,Becker:2011,Viel:2013a,
Lidz:2014,Garzilli:2015,Nasir:2016}.

In the light of the new Planck constraints on reionization, as well as the 
increasing number of quasars discovered at high-$z$ 
\citep[e.g.][]{Banados:2014,Matsuaoka:2016,Banados:2016}, it is pertinent to 
revisit what \lyalpha{} observations with current and upcoming facilities can 
tell us about $\HI$ reionization. This work aims to explore in detail the 
possibilities of using the \lyalpha{} forest 1D flux power spectrum at high-$z$ 
to constrain HI reionization. For this we have used a new method that has recently 
been introduced by \citet{Onorbe:2017}. It builds on the \citet{Haardt:2012} model, 
but enables one to vary the redshifts of $\HI$ and $\HeII$ reionization, as well 
as their associated heat injection, allowing one to consistently simulate a more diverse 
range of reionization histories. This allows for a much comprehensive and 
consistent exploration of the space of thermal parameters than previously done. 
We present here the results of the 1D flux power spectrum at high-$z$ of a new 
set of hydrodynamical simulations that also improve the resolution used in 
previous studies at these redshifts. Additionally, we compare these power spectra
with some recent measurements at these redshifts \citep{Viel:2013a}.

The structure of this paper is as follows. In Section~\ref{sec:sims} we describe 
the characteristics of our hydrodynamical simulations and the different \HI{} 
reionization models studied in this work. Section~\ref{sec:thermal histories} 
presents the thermal evolution of the different reionization models obtained 
from the cosmological hydrodynamical simulations. In Section~\ref{sec:sshydro} 
we compare the 1D flux power spectrum of each model at $5\leq z\leq6$ as well as 
with the best available observations. We discuss in Section~\ref{sec:disc} the 
relevance of our findings in the context of current observational and 
theoretical limitations. We conclude by presenting a summary of our results and 
an outlook in Section~\ref{sec:conc}. In Appendix~\ref{app:convergence}
we perform a set of convergence test for the optical depth and the
1D flux power spectrum at $5\leqslant z\leqslant6$.

Throughout this paper we assumed a flat $\Lambda$CDM cosmology with the following 
fundamental parameters: $\Omega_{\rm m} = 0.3192$, $\Omega_{\rm \Lambda} = 
0.6808$, $\Omega_{\rm b} = 0.04964$, $h = 0.67038$, $\sigma_{\rm 8} = 0.826$ and 
$n_{\rm s} = 0.9655$. These values agree within one sigma with the latest
cosmological parameter constrains from the CMB \citep{Planck:2015,Planck:2016b}. The mass 
abundances of hydrogen and helium ($X_{\rm p} = 0.76$ and $Y_{\rm p} = 0.24$)
were chosen to 
be in agreement with the recent CMB observations and Big Bang nucleosynthesis 
\citep{Coc:2013}.

\section{Simulating Reionization Histories in Light of New Planck Constraints}
\label{sec:sims}

The cosmological hydrodynamical simulations used in this work were performed 
using the Nyx code \citep{Almgren:2013}.  Application of Nyx to studies of the 
\lyalpha{} forest, and its convergence and resolution requirements are discussed 
in \citet{Lukic:2015}. We refer to these two works for more details of the 
numerical implementation, accuracy, and code performance. To generate the 
initial conditions for the simulations, we have used the \textsc{music} code 
\citep{Hahn:2011}, with the transfer function for our cosmological model
obtained from \textsc{camb} 
\citep{Lewis:2000,Howlett:2012}. All simulations discussed in this work used the 
same initial conditions and have a box size of length $L_{\rm box}=20$ Mpc h$^{-1}$ and 
$1024^3$ resolution elements.

As is standard in hydrodynamical simulations that model the \lyalpha{} forest, 
all cells are assumed to be optically thin to radiation. Thus, radiative feedback 
is accounted for via a spatially uniform, but time-varying
ultraviolet background (UVB) radiation field,
input to the code as a list of photoionization and photoheating rates that 
vary with redshift \citep[e.g.][]{Katz:1996}. In order to simulate each 
reionization model discussed here, we have used the method presented in 
\citet{Onorbe:2017}, which allows us to vary the timing and duration of 
reionization, and its associated heat injection, enabling us to simulate a 
diverse range of reionization histories.  This method allows us to create the 
\HI{}, \HeI{} and \HeII{} photoionization and photoheating rates, which are 
inputs to the Nyx code, by volume averaging the photoionization and energy 
equations. This method requires that each reionization event is defined by 
the ionization history with redshift, e.g. $x_{\HI}(z)$, and its associated 
total heat injection, $\Delta T$, which depends on the spectral shape and 
abundance of the ionizing sources, and the opacity of the IGM 
\citep{Abel:1999,McQuinn:2012,Davies:2016,Park:2016}. We direct the reader to 
\citet{Onorbe:2017} for the details of this method.

\begin{figure*}
\begin{center}
\includegraphics[angle=0,width=0.49\textwidth]{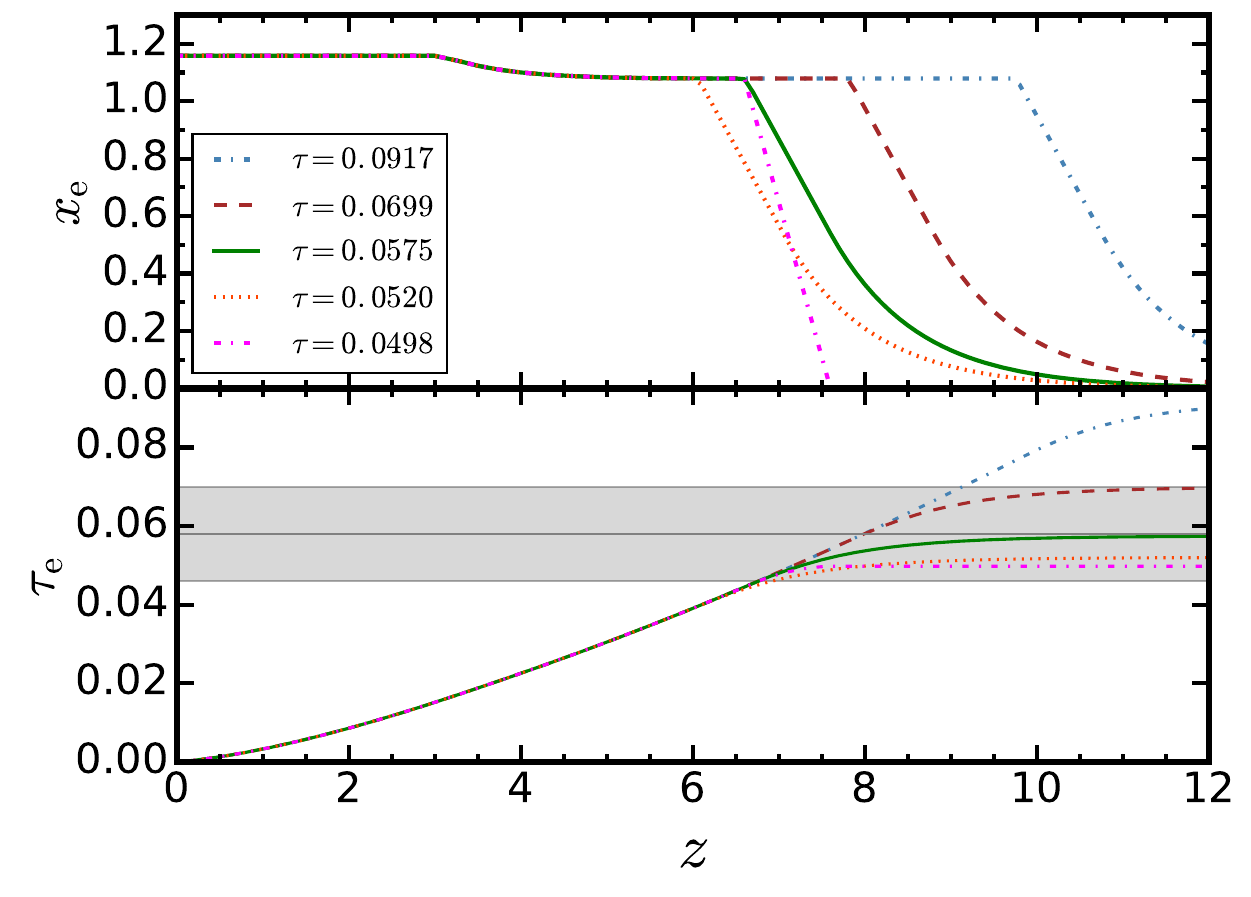}
\includegraphics[angle=0,width=0.49\textwidth]{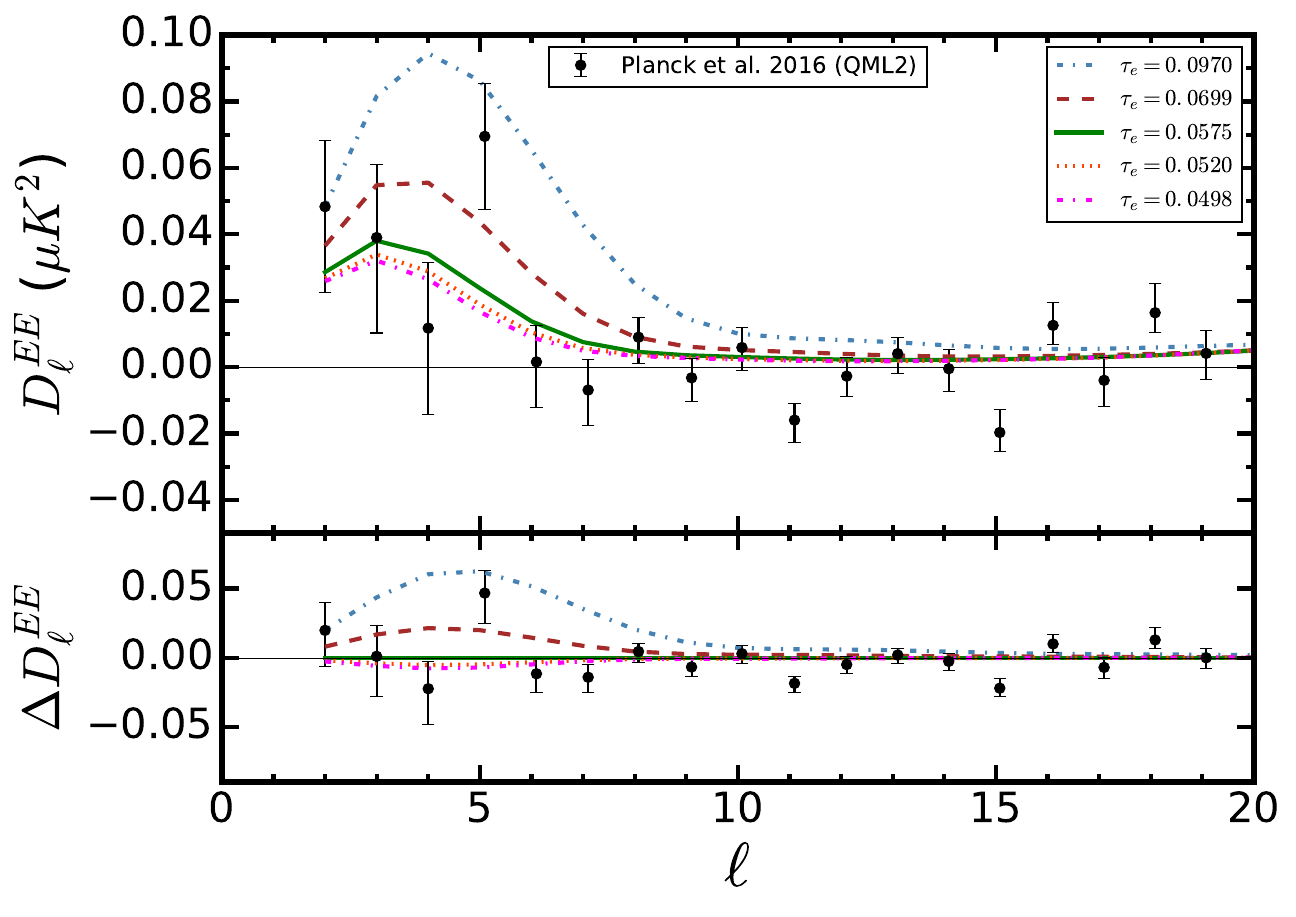}
\end{center}
\caption{Reionization models studied in this work. Upper left panel: evolution
of the free 
electron fraction, $\xe$, for the different reionization models 
considered in this work. Lower left panel: integrated electron-scattering 
optical depth, $\taue$, computed from the above models.  The gray band stands for 
the last constraints on $\taue$ coming from \citet{Planck:2016b} data. 
The right panel shows the low multipole EE power spectrum for the same reionization 
models for which we run hydrodynamical simulations
to compare with the last CMB results from \citet[][black circles,]{Planck:2016a}.
\label{fig:Ionhistory}}
\end{figure*}

In order to determine the reionization histories for our simulations, we explore 
the relevant range of reionization models considering the last $\taue$ 
measurements by Planck and the \lyalpha{} optical depth at high redshift, which 
set a lower limit for $\HI$ reionization at $z=6$ \citep{Fan:2006,McGreer:2015,Becker:2015}.
In all simulations we also assumed the same $\HeII$ reionization model ending at 
$z=3$ and which does not produce a significant increase in the IGM temperature 
until $z<5$. This model is the same as assumed in the \citet{FaucherGiguere:2009} UVB 
model, and since we will be comparing with data at $z\geq5$ this assumption
will not impact our results \citep[see][for more details]{Onorbe:2017}. In this work we 
consider five model $\HI$ reionization histories constructed using the 
analytical formula presented in \citet{Onorbe:2017} chosen to match the
results of radiative  transfer simulations \citep{Pawlik:2009},
\begin{equation} \mean{x_{\HII}}=\begin{cases} 
0.5+0.5\times g(1/n_{1},|z-z_{\rm reion,\HI}^{0.5}|^{n_{1}}), \\ \hspace{3.5cm} 
z <= z_{\rm reion,\HI}^{0.5} \\ 0.5-0.5\times g(1/n_{2},|z-z_{\rm 
reion,\HI}^{0.5}|^{n_{2}}), \\ \hspace{3.5cm} z > z_{\rm reion,\HI}^{0.5} \\ 
\end{cases} \label{eq:Qana} \end{equation} where $g$ is the incomplete gamma 
function, $n_{1}=50$, $n_{2}=1$ and $z_{\rm reion,\HI}^{0.5}$ is a free 
parameter that sets the redshift where $x_{\HII}(z_0)=0.5$.

We run an early, middle, and late $\HI$ reionization history (EarlyR, MiddleR, 
LateR), which have a specific reionization redshifts (defined as the redshift 
where $\mean{x_{\HII}}=0.999$) of $z_{\rm reion,\HI}= 7.75$, $6.55$, and $6.0$ 
respectively, 
and are within $1\sigma$ of the Planck CMB measurements. We also run two more 
models, one with a very early reionization (VeryEarlyR, $z_{\rm reion,\HI}= 
9.70$ which is $3\sigma$ discrepant with the Planck measurement)
and a faster reionization (MiddleR-fast, $z_{\rm reion,\HI}= 6.55$). A 
summary of all the relevant parameters used in the runs presented in this work 
is shown in Table~\ref{tab:sims} along with the naming conventions we have 
adopted.

\begin{table*}
\begin{center}
\caption{Summary of Simulations. \label{tab:sims}}
\begin{footnotesize}
\begin{tabular}{lcccccccc}
\tableline\tableline
Sim & $\HI$ reionization & $z_{\rm reion,\HI}$ & $z_{\rm reion,\HI}^{0.5}$ & $\Delta z$  & $\tau_{\rm e}$& $\Delta T_{\HI}$ & $u_{0}(z=4.9)$\\
     &  & & & & &(K) & (eV $m_{\rm p}^{-1}$)  \\
\tableline
VeryEarlyR     & Very Early  & $9.70$ & $10.75$ & $2.59$ & $0.0917$ & $2\times10^{4}$&$7.86$\\ 
EarlyR         & Early       & $7.75$ & $8.80$ & $2.59$ & $0.0698$ & $2\times10^{4}$ &$5.52$\\ 
MiddleR        & Middle      & $6.55$ & $7.60$ & $2.59$ & $0.0574$ & $2\times10^{4}$ &$4.24$\\ 
LateR          & Late        & $6.00$ & $7.05$ & $2.59$ & $0.0520$ & $2\times10^{4}$ &$3.85$\\ 
MiddleR-fast   & Fast middle & $6.60$ & $7.10$ & $0.89$ & $0.0498$ & $2\times10^{4}$ &$4.18$\\ 
MiddleR-cold   & Middle      & $6.55$ & $7.60$ & $2.59$ & $0.0574$ & $1\times10^{4}$ &$3.16$\\ 
MiddleR-warm   & Middle      & $6.55$ & $7.60$ & $2.59$ & $0.0574$ & $3\times10^{4}$ &$5.60$\\ 
MiddleR-hot    & Middle      & $6.55$ & $7.60$ & $2.59$ & $0.0574$ & $4\times10^{4}$ &$7.08$\\ 
\tableline
\end{tabular}
\end{footnotesize}
\tablecomments{All simulations have a box size of length $L_{\rm box}=20$ 
Mpc h$^{-1}$ and $1024^3$ resolution elements. 
Column 1: Simulation code.
Column 2: $\HI$ ionization history assumed for the model.
Column 3: $\HI$ reionization redshift. Redshift where $x_{\HII}(z_0)=0.999$.
Column 4: redshift at which $\HI$ reionization is halfway.
Column 5: Width of $\HI$ reionization. $\Delta z=z_{\rm reion,\HI}^{0.99}-z_{\rm reion,\HI}^{0.1}$.
Column 6: CMB integrated electron-scattering optical depth.
Column 7: Total heat input assumed for $\HI$ reionization used to build the UVB models.
Column 8: Total energy per particle injected during $\HI$ reionization.
Column 9: Cumulative energy deposited parameter defined by \citet{Nasir:2016} at $z=4.9$.
See text for more details.}
\end{center}
\end{table*}

\begin{figure*}
\begin{center}
\includegraphics[angle=0,width=0.49\textwidth]{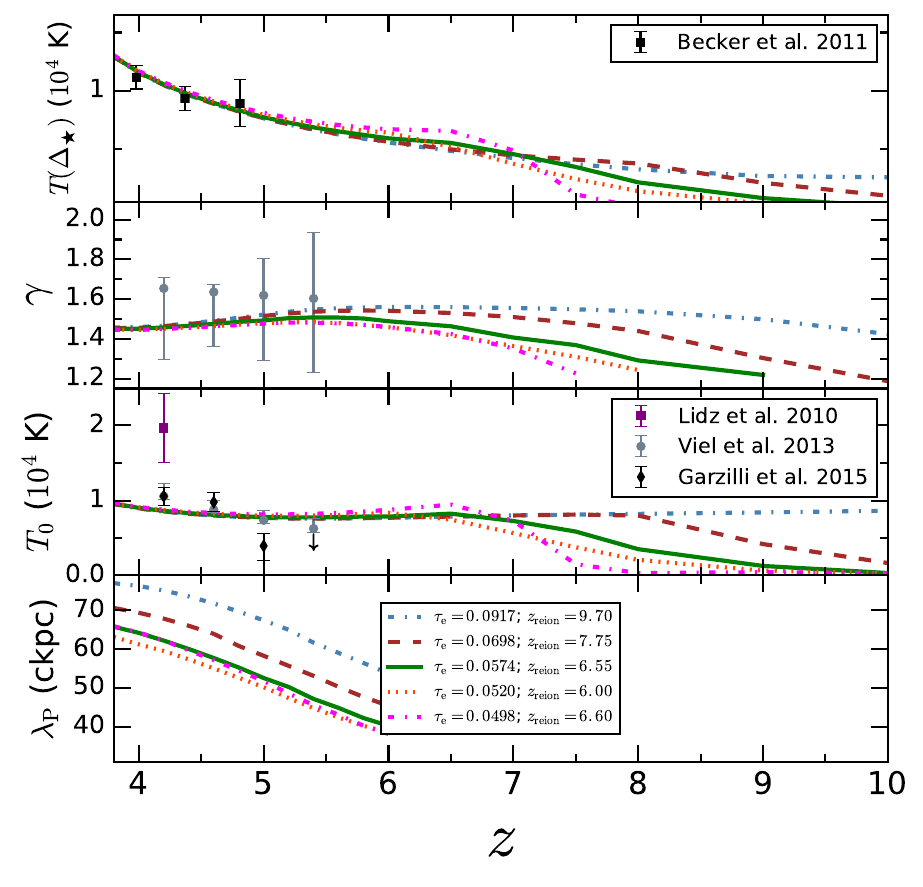}
\includegraphics[angle=0,width=0.49\textwidth]{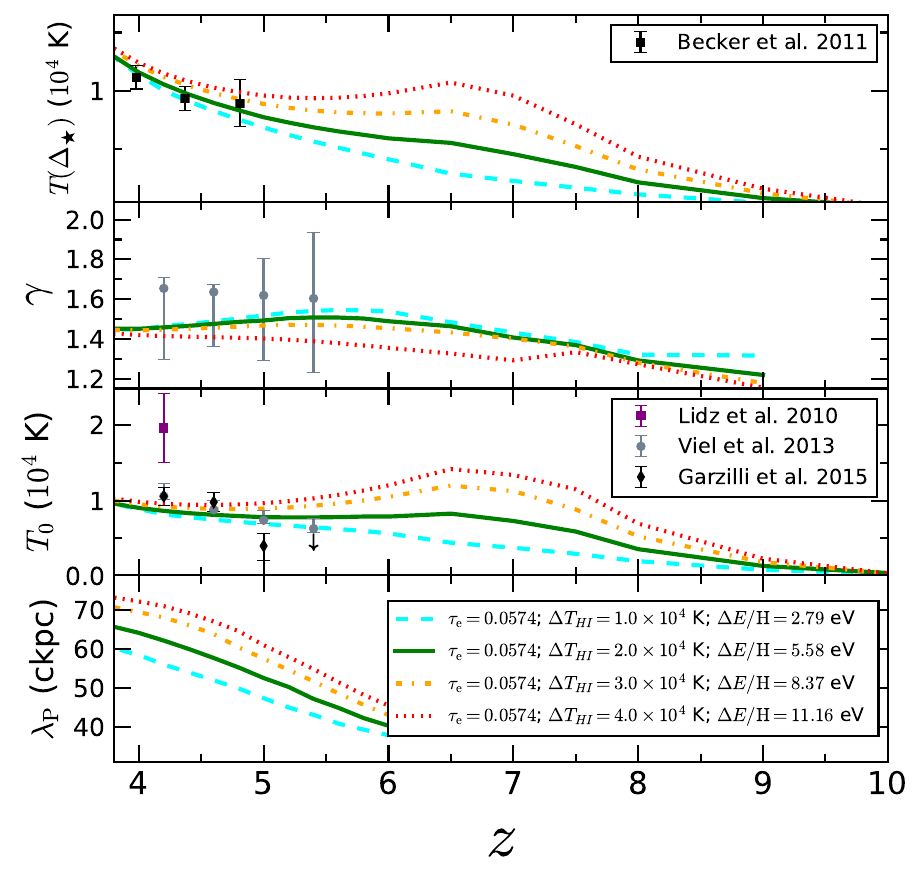}
\end{center}
\caption{
Thermal history obtained in simulations using different UVB models. The left panels 
present the thermal history of the simulations in which we changed when $\HI$ 
reionization happened, but used the same heat input during reionization. The 
thermal history of the simulations in which $\HI$ reionization happened at the 
same time, but the input heat was changed appears on the right panels. In both 
cases they display the evolution of the different thermal parameters in
the simulation: the temperature at the optimal 
density, $T(\Deltaopt)$ (top), the slope of 
the density-temperature relation, $\gamma$ (second top), the 
temperature at mean density, $T_0$ (second bottom), the pressure-smoothing scale, $\lj$, (bottom).
Note that while the temperature is just sensitive 
to the current photoionization and photoheating values, the actual pressure 
smoothing scale value depends on the full thermal history of each simulation. Symbols 
with error bars stand for different observational measurements and their 
$1\sigma$ error. See text for more details.
\label{fig:thermalhistory}}
\end{figure*}

The full reionization history of each of our models is plotted in the upper left 
panel of Figure~\ref{fig:Ionhistory}, shown as the redshift evolution of the 
electron fraction given by $\xe=\nel/\nH=(1+\chi)\mean{x_{\HII}} +\chi 
\mean{x_{\HeIII}}$ where $\chi=Y_{p}/(4X_{p})$ and $X_{p}$ and $Y_{p}$ are the 
hydrogen and helium mass abundances, $x_{\HII}(z)$ is the hydrogen ionized 
fraction, and $x_{\HeIII}(z)$ is the fraction of helium that is doubly 
ionized\footnote{Throughout this paper we made the standard assumption that 
$\HeI$ reionization is perfectly coupled with that of $\HI$.}. The lower left 
panel of Figure~\ref{fig:Ionhistory} shows the evolution of the 
cosmic reionization optical depth, $\taue$, for each of these models. 
In the right panel of Figure~\ref{fig:Ionhistory} we compare the newest 
Planck measurements of the CMB polarization EE power 
spectrum\footnote{Throughout this paper we adopt the convention 
$D_{\ell}=\ell(\ell+1)C_{\ell}/2\pi$.} low multipoles \citep[][]{Planck:2016a} 
to these reionization models, where we have computed the EE power spectrum using 
the \textsc{class} Boltzmann code \citep{Lesgourgues:2011}.  For the observed 
low multipoles we show the unbiased QML$_{2}$ results from \citet[][black 
circles]{Planck:2016a}, illustrating the impressively high precision achieved by 
these CMB polarization measurements, which significantly reduces the allowed 
range of models.

To build the reionization models, we also need to assume the associated total 
heat injection, $\Delta T$, during $\HI$ reionization, which depends on the 
spectral shape and abundance of the ionizing sources, and the opacity of the IGM 
\citep{Abel:1999,Tittley:2007,McQuinn:2012,Davies:2016,Park:2016}.
To run all 
the simulations with the reionization model described above, we assumed $\Delta 
T_{\HI} = 2 \times10^{4}$ K which is the standard value obtained in 
galaxy-driven $\HI$ reionization models using 1D radiative transfer simulations
\citep[e.g.][]{McQuinn:2012}. Quasar-driven scenarios give higher heat injection
values, $\Delta T_{\HI} \sim 4 \times10^{4}$.
Thus, in order to study the effect of different total heat input during 
$\HI$ reionization, we run three more simulations with the same $\HI$ 
reionization model as MiddleR, but varying the $\Delta T_{\HI}$ parameter: 
MiddleR-cold ($\Delta T_{\HI} = 1 \times10^{4}$ K), MiddleR-warm ($\Delta 
T_{\HI} = 3 \times10^{4}$ K), and MiddleR-hot ($\Delta T_{\HI} = 4 \times10^{4}$ 
K).

\section{Reionization-dependent Thermal Histories}
\label{sec:thermal histories}

Changing the timing and duration of reionization and its associated heat 
injection will manifest as changes in the evolution of the parameters governing 
the thermal state of the IGM. In Figure~\ref{fig:thermalhistory} we present the 
resulting thermal histories for all of these simulations. The upper panel shows 
the evolution of temperature $T(\Deltaopt)$ at the `optimal' overdensity 
$\Deltaopt$ probed by curvature measurements of the \lyalpha{} forest 
\citep[see][]{Becker:2011,Boera:2014}, where we calculate the optimal density at 
each redshift using the functional form of $\Deltaopt(z)$ given by 
\citet{Becker:2011}. The evolution of thermal parameters, $\gamma$ and $T_{0}$, 
governing the density-temperature
relation, are shown in the second and third 
panel from the top, determined by fitting the distribution of densities and 
temperatures in the simulation following the linear least-squares method 
described in \citet{Lukic:2015}\footnote{ Changing the thresholds used to do the 
fit within reasonable IGM densities produces differences just at a few percent 
level \citep[see][for similar conclusions]{Lukic:2015} and in any case it does 
not affect the conclusions presented in this work. We also found no relevant 
effects in the main results of this paper when we employed the fitting 
approach used in \citet{Puchwein:2015}.}.

The evolution of the pressure-smoothing scale, $\lj$, with redshift is shown in 
the bottom panel. To characterize the pressure-smoothing scale in all our 
simulations we have followed the approach described by \citet{Kulkarni:2015}. 
These authors define a pseudo real-space \lyalpha{} flux field, which is the 
same as the true \lyalpha{} forest flux, but without redshift space effects such 
as peculiar velocities and thermal Doppler broadening. This field naturally 
suppresses the dense gas that would otherwise dominate the baryon power spectrum, 
making it robust against the poorly understood physics of galaxy formation
and revealing the pressure-smoothing in the diffuse IGM.

Inspection of the left panel of Figure~\ref{fig:thermalhistory} reveals that 
simulations with different reionization histories but the same heat
injection during reionization, $\Delta T$, all 
share a very similar $T_{0}$, $T(\Deltaopt)$ and $\gamma$ evolution at $z=5-6$.
This is because once reionization is completed, the IGM thermal state
asymptotes to a tight power-law temperature-density relation
driven mainly by the photoheating rate and accelerated by
Compton and adiabatic cooling.
The time to converge to these asymptote values 
is around $\Delta z\sim 1-2$ (a few hundred Myr) 
once reionization is completed and mainly depends
on the amount of heat injected during reionization \citep[][]{McQuinn:2016}.
Since all these models
share the same photoionization and photoheating values, once reionization
is completed they all converge to the same thermal state 
at lower redshift $z\lesssim6$.
However, their pressure-smoothing scale, $\lj$, remains 
very different at these and lower redshifts. Models in which reionization 
happened at earlier times have a larger pressure-smoothing scale. As discussed 
above, this results from the dependence of the IGM pressure-smoothing scale on 
the full thermal history \citep{Hui:2003, Kulkarni:2015, Onorbe:2017} and not 
just on the instantaneous temperature, and it will have 
important consequences for the statistics of the \lyalpha{} forest 
\citep[][]{Nasir:2016,Onorbe:2017}.

The right panels of Figure~\ref{fig:thermalhistory} show the thermal histories 
for simulations where we fixed the reionization history, but varied the total 
heat injection during $\HI$ reionization. Models with more heat injection give 
rise to higher temperatures and slightly lower $\gamma$ values not only during 
reionization, but also at later times while the IGM is still 
reaching their asymptote values.
Note that similarly as for the different reionization history models, these models 
also produce a larger pressure-smoothing scale, and these differences persist even 
at lower redshifts long after the other thermal parameters, $\gamma$ and $T_{0}$, 
have relaxed to their asymptotic values.
Once 
reionization is completed all these models share the same photoionization and 
photoheating rates and therefore $\gamma$, $T_{0}$, and $T(\Deltaopt)$ thermal 
parameters asymptote to the same values much faster than the pressure-smoothing 
scale, which retains memory of the thermal history.

The symbols with error bars in Figure~\ref{fig:thermalhistory} indicate recent 
observational constraints on the parameters governing the thermal state of the 
IGM at high redshift. In particular, the purple square is the \citet{Lidz:2010} 
measurement of $T_{0}$ using wavelets at $z = 4.20$ and black squares are the 
\citet{Becker:2011} measurements of $T(\Deltaopt)$ based on the curvature 
statistic. Gray circles and black diamonds represent the joint fits to $\gamma$ 
and $T_{0}$ given by \citet{Viel:2013a} and \citet{Garzilli:2015} respectively, 
using the 1D flux power spectrum.\footnote{These measurements are marginalized 
  over the mass of a warm dark matter particle.
\citet{Viel:2013a} and \citet{Garzilli:2015} used different
fitting approaches, but both used the 
same grid of hydrodynamical simulations in which the standard reionization 
redshift for the runs was $z_{\rm reion}=12$ and the lowest reionization 
redshift considered in the grid, by including one simulation, was $z_{\rm 
reion}=8$.}
While the thermal parameters measured 
by \citet{Becker:2011}, \citet{Viel:2013a} and 
\citet{Garzilli:2015}
appear consistent with the 
models discussed here, the \citet{Lidz:2010} $T_0$ measurement suggests a 
significantly hotter IGM at $z\sim4$.
The origin of this disagreement is unclear, but may 
result from differences in the methods used by these authors and/or the 
different hydrodynamical simulations compared to the data. Our work aims to 
shed more light on this issue by comparing
the 1D power spectrum measurements at high 
redshift with an improved set of simulations.

\section{The 1D Flux Power Spectrum at High Redshift $z \sim 5-6$}
\label{sec:sshydro}

\begin{figure*}
\begin{center}
 \includegraphics[angle=0,width=0.48\textwidth]{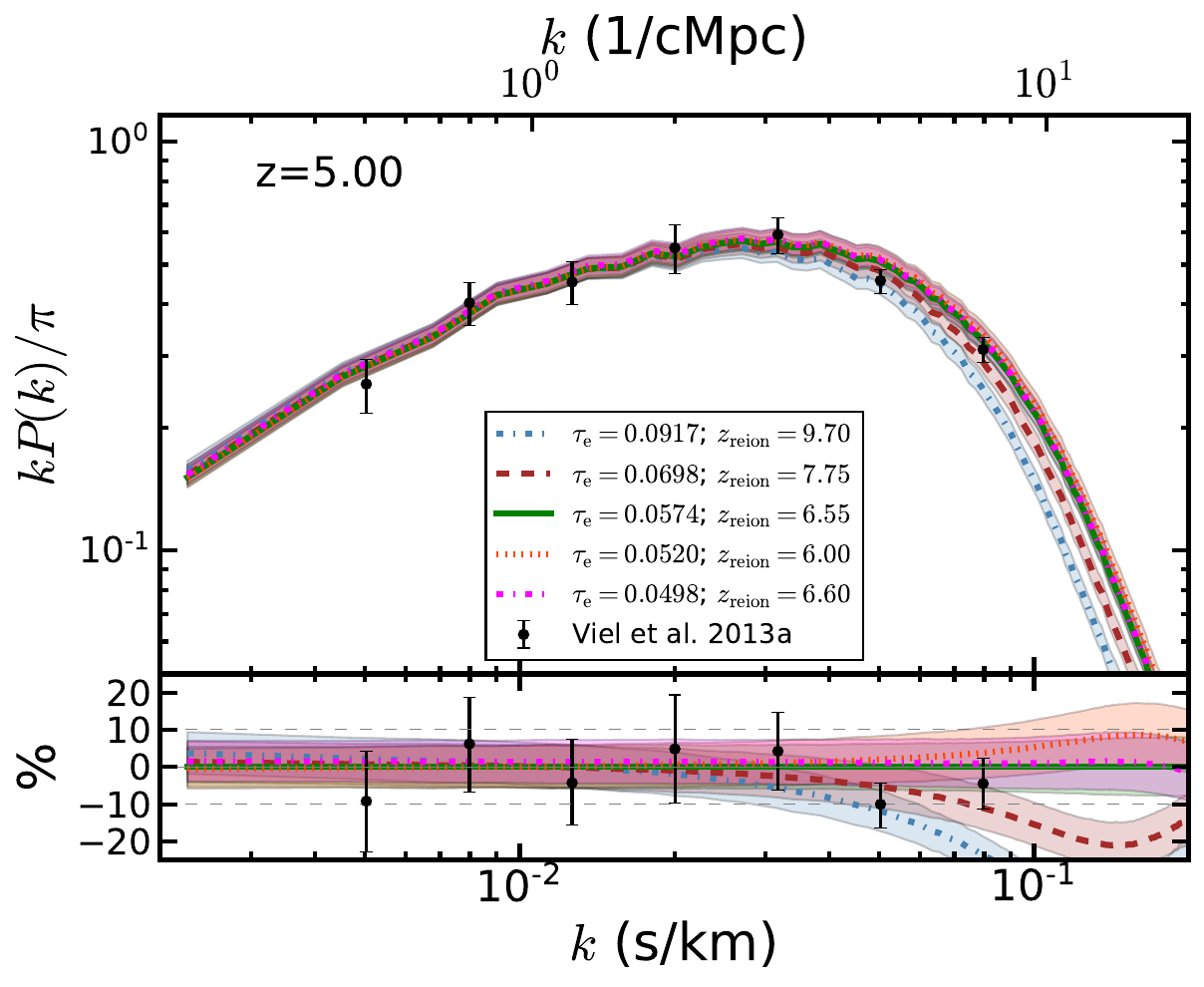}
 \includegraphics[angle=0,width=0.48\textwidth]{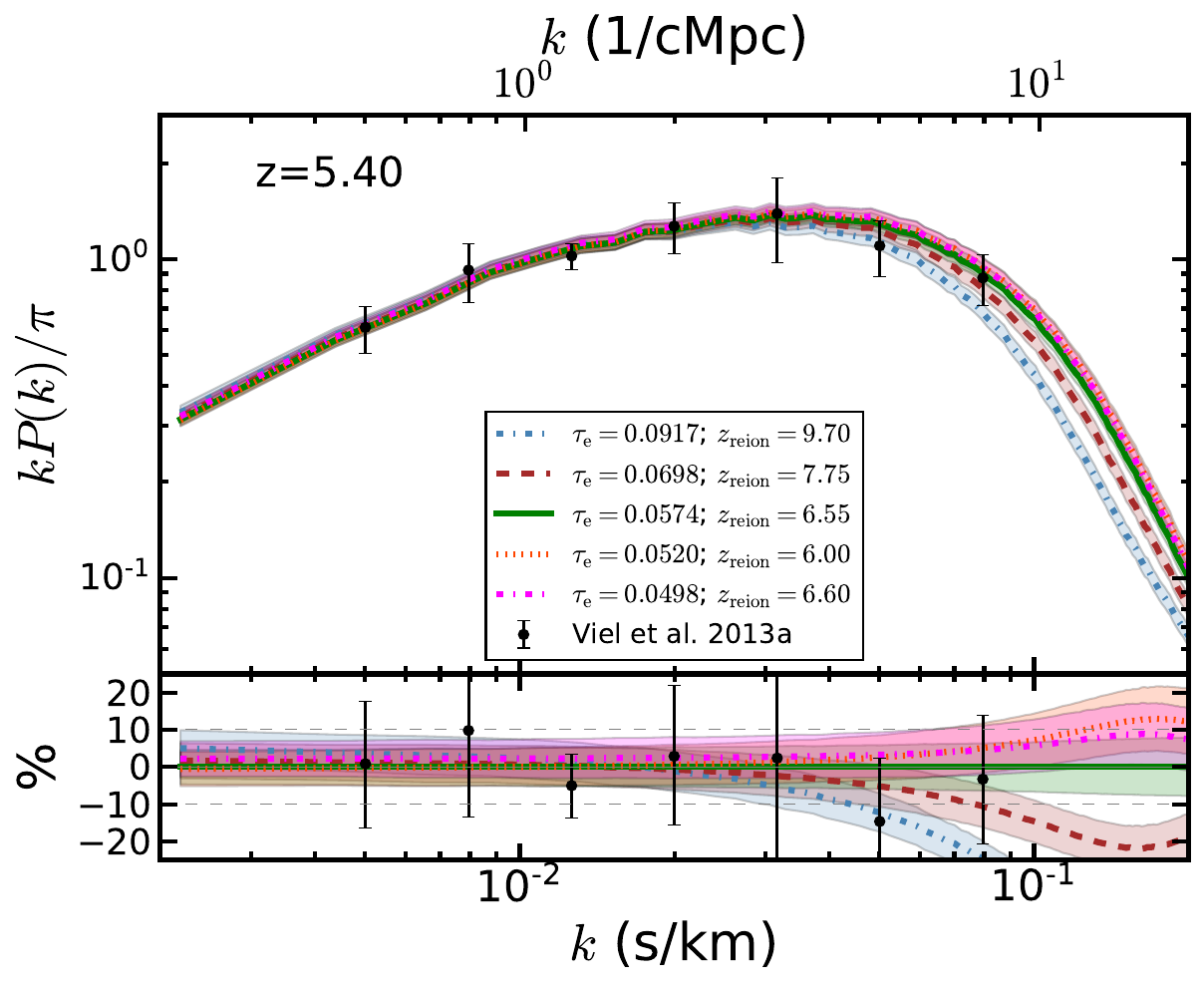}
\end{center}
\caption{Effect of a different $\HI$ reionization history on the 1D flux power spectrum at 
$z=5.0$ and $z=5.4$. Simulations that differ in their 
$\HI$ reionization history: EarlyR, MiddleR, LateR, MiddleR-fast and VeryEarlyR.
Notice that at $z=5$ these simulations have a very similar IGM temperature 
$T_{0}$ but mainly differ in their pressure-smoothing scale. 
Black circles in $z=5$ plots 
stand for observational measurements done by \citet{Viel:2013a} using 
high-resolution spectra of 25 quasars with emission redshifts $4.48 \leq z_{em} 
\leq6.42$. 
Color bands show the variation in the 1D flux power spectrum due to
one sigma changes in the mean flux at the corresponding redshift.
See text for more details.
\label{fig:ps1dzreion}}
\end{figure*}

In order to explore the possibility of discriminating between the reionization 
models presented here with \lyalpha{} forest measurements, we calculate the 1D 
flux power spectrum, $P(k)$, for each simulation at $z=5.0$, $5.4$, and $6.0$.  The 
1D power spectrum of the \lyalpha{} forest is sensitive to the parameters 
governing the thermal state of the IGM. Pressure smoothing, $\lambda_{\rm P}$, 
damps out small-scale fluctuations in the gas, while random thermal motions 
(sensitive to temperature, or $T_0$ and $\gamma$) Doppler broadens \lyalpha{} 
forest lines, further reducing the amount of small-scale structure. Both of 
these effects combine to produce a prominent small-scale (high-k) cutoff in the 
flux power spectrum $P(k)$ \citep{Zaldarriaga:2001, 
Peeples:2010a,Rorai:2013,Nasir:2016}.  Therefore, by carefully studying this 
cutoff, we expect to be able not only to constrain the thermal state of the gas,
but also its full thermal history.

We have created \lyalpha{} forest spectra from the simulation computing the 
$\HI$ optical depth at a fixed redshift, which can then be easily converted into 
a transmitted flux fraction, $F_{\HI} = e^{−\tau_{\HI}}$. We refer to 
\citet{Lukic:2015} for specific details of these calculations. We computed the 
power spectrum, $P(k)$, of the fractional contrast,
$\delta F$, at each 
redshift defined as $\delta F$=$F/\mean{F} -1$. A total of $1024^2$ skewers were
used at each redshift. We computed the power spectrum of each skewer and 
then calculated the average value at each mode, $k$.

The overall level and precise shape of the fractional contrast power is still
sensitive to the mean flux because it changes the density-flux mapping.
This is to say, a lower mean flux will shift \lyalpha{} observations sensitivity
toward lower densities. For this reason, when computing the 1D power spectrum 
of the fractional contrast from a simulation, it is still very important
which mean flux was considered.
Following the standard approach, we rescaled the mean flux of
each simulation to match a fixed mean flux value.
Of course, this rescaling does not affect measurements 
since one directly measures a flux contrast field.
While the mean flux value has been precisely measured at lower
redshift, the measurements at $z\gtrsim 5$ are more uncertain.
At $z=5.0$ and $5.4$, the current best measurements
for the mean flux are the binned 
values computed by \citet{DAloisio:2016} from the \citet{Becker:2015} high-redshift
quasar opacity measurements.
These are $\mean{F}=0.14\pm0.01$ for 
$z=5$, and $\mean{F}=0.08\pm0.006$ for $z=5.4$.
These values are consistent with 
the analytic formula presented by \citet[]{Viel:2013a}
derived from their own quasar sample, which are 
$\mean{F}=0.14603,0.071$ respectively. The
\citet[]{Fan:2006} measurements of the 
mean flux using a sample of high-$z$
quasars discovered in the Sloan Digital Sky 
Survey were $\mean{F}=0.1224\pm0.03$ at $z=5.025$ and 
$\mean{F}=0.074^{+0.03}_{-0.06}$ at $z=5.450$. 
These values are also in good agreement with the current best
measurements considering their larger errors.
In fact, the global fit suggested by \citet{Fan:2006} 
based on their own measurements
gives $\mean{F}=0.1659,0.071$, respectively at
these redshifts.
However, since we will compare our models with the \citet{Viel:2013a} observations of 
the 1D flux power spectrum in order to give a better qualitative idea of the 
results found in this work, we have considered the mean mean flux values that 
give a better overall normalization to these observations and assumed 
$\sim7.5$\% relative measurement error, which reflects the quoted errors in the 
results above: $\mean{F}=0.16\pm0.01$ for $z=5$,
and 
$\mean{F}=0.055\pm0.004$ for $z=5.4$. While these values seem to be slightly far 
from the current best observations by \citet{DAloisio:2016} at $z=5.4$,
they are within the $1\sigma$ C.L. found by 
\citet[][see Table II]{Viel:2013a} when they performed a marginalized fit of the 1D 
flux power spectrum for several parameters that included the mean flux 
($\mean{F}=0.148^{+0.024}_{-0.007}$ at $z=5$, and 
$\mean{F}=0.045^{+0.02}_{-0.001}$ at $z=5.4$)\footnote{This is also in agreement 
with the marginalized fit made by \citet[][see Table I]{Garzilli:2015} to the 
same dataset. These authors found $\mean{F}=0.142^{+0.023}_{-0.017}$ at $z=5$ and 
$\mean{F}=0.054^{+0.014}_{-0.01}$ at $z=5.4$.}.
Although the lower mean flux measurements from the 1D flux power spectrum
could just indicate 
some fluctuation due to the small number of quasars used to compute the 
power spectrum at high-$z$, it definitely highlights the relevance of taking into 
account the mean flux degeneracy when
any astrophysical or cosmological 
information from the 1D flux power spectrum is to be extracted. We return
to this issue in our discussion of different degeneracies of
the 1D flux power spectrum in Section~\ref{sec:disc}.

\begin{figure*}
\begin{center}
 \includegraphics[angle=0,width=0.48\textwidth]{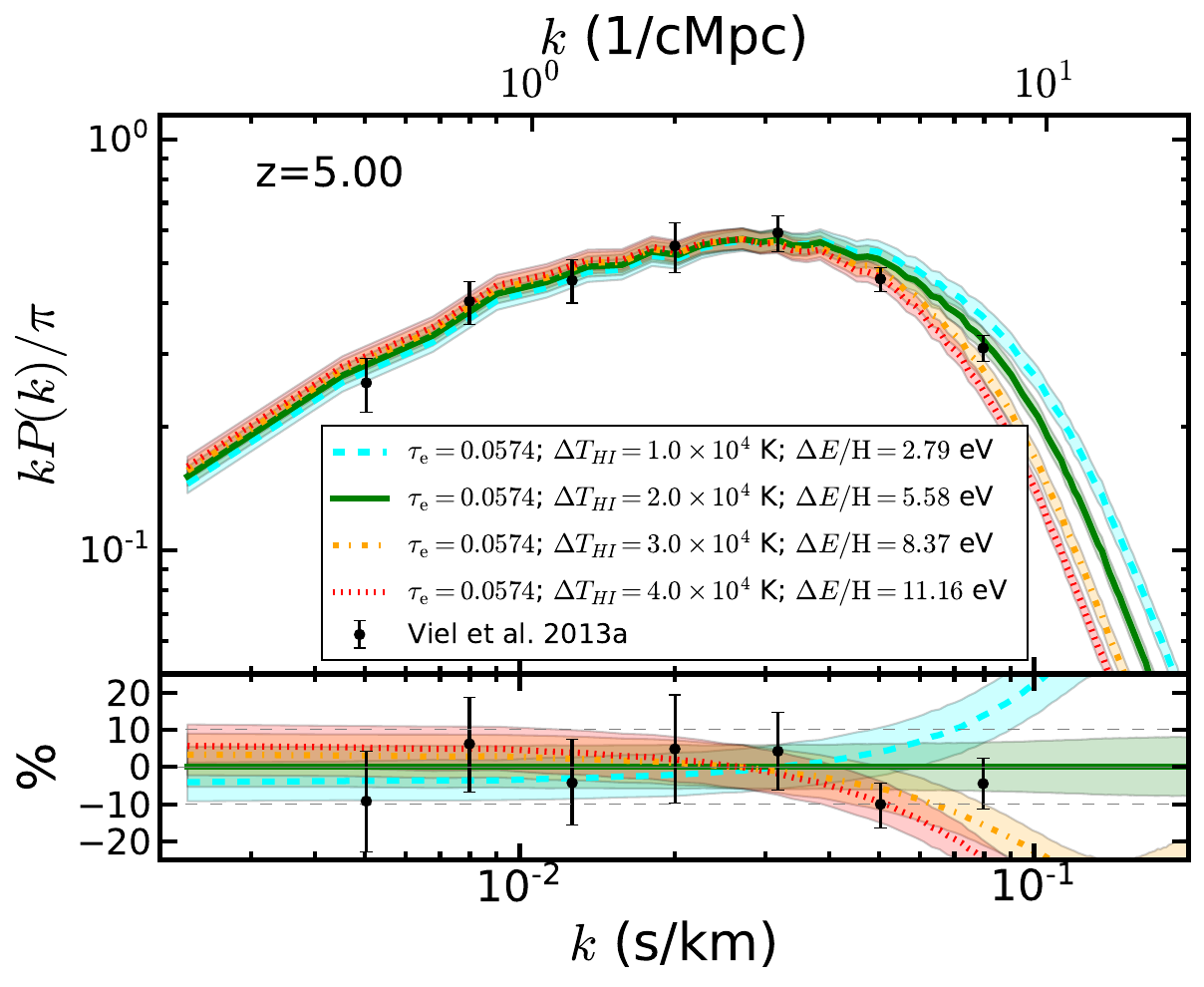}
 \includegraphics[angle=0,width=0.48\textwidth]{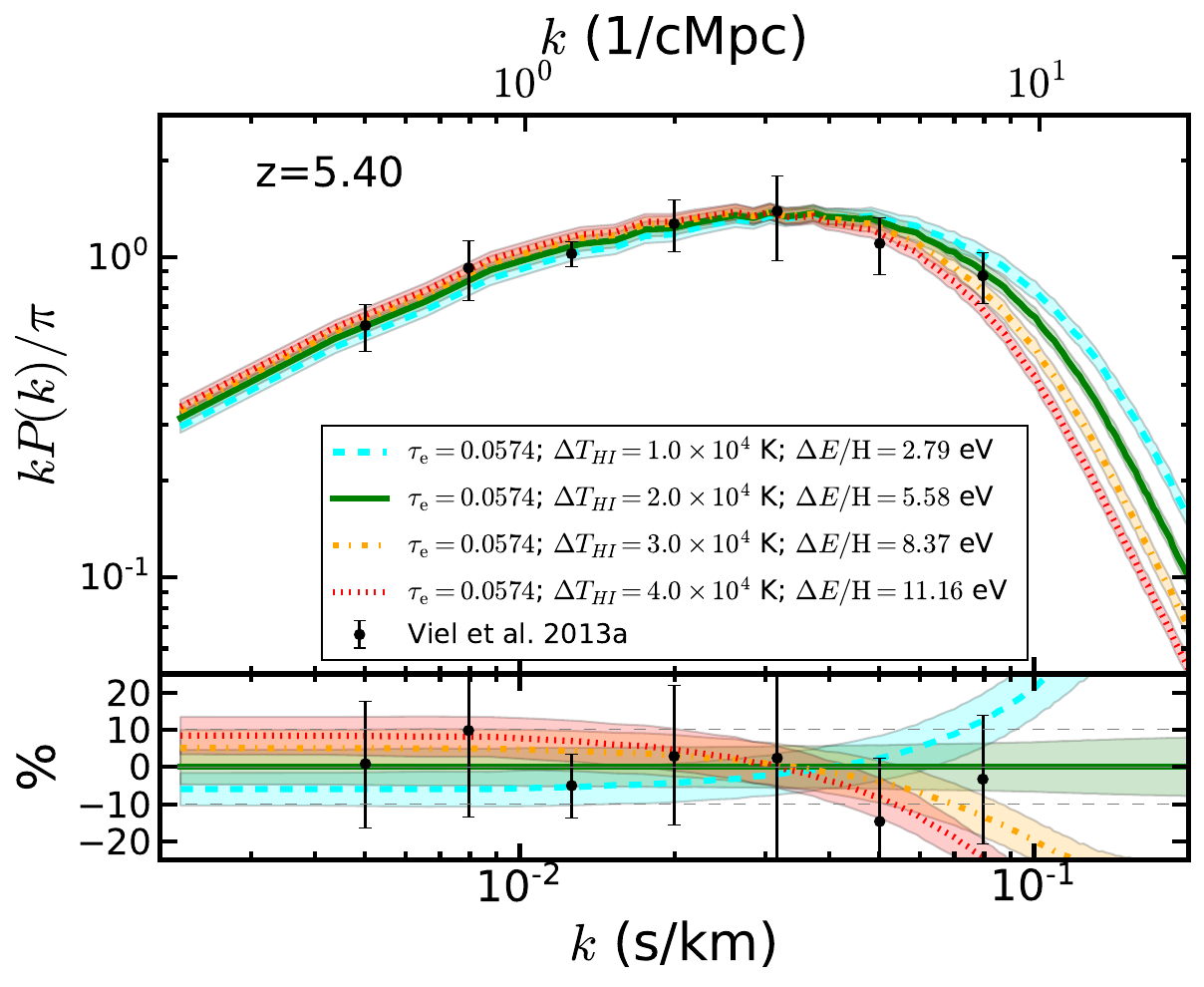}
\end{center}
\caption{Effect of different heat injection during $\HI$ reionization 
on the 1D flux power spectrum at $z=5.0$ and $z=5.4$. 
Simulations in which reionization happened at 
the same time but that differ in the heat input during $\HI$ reionization: 
Middle-cold, MiddleR, MiddleR-warm, MiddleR-hot.
Black circles in $z=5$ plots 
stand for observational measurements made by \citet{Viel:2013a} using 
high-resolution spectra of 25 quasars with emission redshifts $4.48 \leq z_{em} 
\leq6.42$. 
Color bands show the variation in the 1D flux power spectrum due to
one-sigma changes in the mean flux at the corresponding redshift.
See text for more details.
\label{fig:ps1dheat}}
\end{figure*}

The two panels of Figure~\ref{fig:ps1dzreion} show the simulated
dimensionless 1D flux  power spectrum, $kP(k)\slash \pi$, computed at $z=5.0$ and $5.4$
for the models where we changed the $\HI$ reionization history but 
kept the total heat input constant: EarlyR, MiddleR, LateR, MiddleR-fast and 
VeryEarlyR (see Table~\ref{tab:sims}).
We first note in the two panels that the 
overall power level increases with redshift, which reflects the fact that as the 
average mean flux decreases toward higher $z$, density fluctuations are 
exponentially amplified 
\citep[e.g.][]{Viel:2004,PalanqueDelabrouille:2013,Viel:2013a}. Note also that 
at all redshifts the difference between models for the low-$k$ modes is very small and 
therefore the use of high-resolution spectra probing to $k\sim 0.1$ s km$^{-1}$ is 
key.\footnote{Detailed studies of the impact of metal 
absorption features in current state-of-the-art high-resolution spectra have 
shown that it can increase the 1D flux power spectrum at k$>0.1$ s km$^{-1}$
\citep{McDonald:2000,McDonald:2005,Lidz:2010,Viel:2013a}. For this reason, this 
is typically the maximum k considered in high-resolution power spectra studies.}
The models separate at high-$k$ because their disparate reionization 
histories result in different levels of pressure-smoothing and thermal 
broadening (see Fig.~\ref{fig:thermalhistory}), changing the shape of the 
small-scale (high-$k$) cutoffs in the power spectra.
The color bands for each 
model show the variation in the 1D flux power spectrum due to the one-sigma 
uncertainty in 
the mean flux value at the corresponding redshift. 
Note that currently these errors 
seems to translate into a $\sim 10\%$ scatter in the 1D flux power spectrum,
which is smaller than the current error bars on the $z=5.4$ measurements
\citep{Viel:2013b}.

The differences between these models at $z\gtrsim5$ are particularly interesting 
because they result primarily from differences in the pressure-smoothing scale, 
$\lambda_{\rm P}$, as the other parameters governing the thermal state of the 
IGM, $\gamma$ and $T_{0}$, are very similar 
(see left panel of Figure~\ref{fig:thermalhistory}).
These models share exactly the same 
photoionization and photoheating rates at the redshifts considered,  and differ 
solely in the timing of $\HI$ reionization heat injection. These results 
highlight that the 1D \lyalpha{} forest power spectrum is sensitive to the 
details of $\HI$ reionization history even at lower redshifts due to the 
different pressure-smoothing scale. Any attempts to derive astrophysical or 
cosmological parameters using high-$z$ \lyalpha{} forest observations that do 
not take this issue into account could obtain biased results \citep[see 
also][]{Puchwein:2015,Onorbe:2017}.

Figure~\ref{fig:ps1dheat} show the simulated
dimensionless 1D flux  power spectrum for simulations 
in which the timing of $\HI$ reionization is 
identical, but which have different amounts of heat injection, $\Delta T$: 
Middle-cold ($\Delta T=1\times10^4$ K), MiddleR ($\Delta T=2\times10^4$ K), 
MiddleR-warm ($\Delta T=3\times10^4$ K), MiddleR-hot ($\Delta T=4\times10^4$ K). 
As expected, the power spectrum shows a larger
small-scale cutoff (i.e. toward lower $k$) for 
simulations with a higher heat input during reionization. It is clear that the 
effect in the 1D flux power spectrum  of a high heat injection during $\HI$ 
reionization (MiddleR-hot) is degenerate with a reionization model 
with a lower heat input but that completes at higher redshift. Both physical 
processes produce a higher pressure-smoothing scale at lower redshift.

At higher redshifts, $z\gtrsim 5$ the differences in the power spectrum
between simulations shown in Figure~\ref{fig:ps1dheat}
are due not only to the effect of the pressure-smoothing scale, $\lj$, but also 
due to the differences in the other thermal parameters, $\gamma$ 
and $T_{0}$. This is because at these redshifts the IGM in these models
are still reaching the asymptotic temperature-density relation 
after $\HI$ reionization (see Figure~\ref{fig:thermalhistory}).
This highlights the other physical process affecting the \lyalpha{}
forest lines 
which is the thermal broadening along the line of sight that also affects the 
cutoff, in the 1D flux power spectrum. In fact, it is relevant to point out that 
the differences between models with different $T(\Deltaopt)$ produce larger differences 
at $k<0.04$ s km$^{-1}$ in the 1D flux power spectrum 
than thermal models that just differ in the pressure-smoothing scale, $\lj$ (see 
Figure~\ref{fig:ps1dheat}).
This could open a possibility to distinguish between 
both physical effects using different $k$-mode 
ranges of the 1D flux power spectrum, provided that $\HI$
reionization happens at low enough redshift to still see these effects.

\subsection{Comparison with Observations}
\label{ssec:simsvsobs}

\citet{Viel:2013a} made measurements of the 1D \lyalpha{} forest 
flux power spectrum at $z=5.0$ and $z=5.4$,
using a sample of 25 high-resolution quasar spectra. The redshift bins had width
$dz = 0.4$ and contained data from $\sim10$ quasars per bin. 
In Figure~\ref{fig:ps1dzreion} and Figure~\ref{fig:ps1dheat} we also
compare the 
measurements of \citet[][black circles]{Viel:2013a} to our simulation results
at the same redshift bins.  
From these figures it is clear that these measurements already have sufficient 
precision to begin distinguishing between different reionization models, once 
the degeneracy due to the mean flux has been taken into account. In what follows 
we report a first qualitative comparison of this data set with our simulations. 
A detailed quantitative analysis of these observations using a larger hydrodynamical 
grid of high-resolution large-volume simulations that expands the full parameter 
space of the thermal parameters and takes into account relevant degeneracies 
will be presented in another paper.

We first note from this comparison is that in the context of 
our current models and with the caveat that the mean fluxes have been chosen to best fit
the power at each redshift,
the $z=5$ and $z=5.4$ measurements appear to be in agreement with 
our fiducial model (MiddleR, green line) that uses the Planck $\tau_{e}$ value 
and $\Delta T= 2\times10^{4}$ K. This picture is consistent with
the conventional 
wisdom that galaxies reionized hydrogen \citep{Robertson:2015}.
Reionization models driven by active galactic 
nuclei \citep[AGN; see e.g.][]{Chardin:2015,Madau:2015,Khaire:2016}
have recently gained traction in light of the 
discovery of an abundant population of faint AGN at high redshift $z\sim4-6$ 
\citep{Giallongo:2015}. Such models have higher photoelectric heating 
of $\HI$ and would also doubly ionize helium at
these high redshifts \citep{McQuinn:2012} increasing the amount of heat injection 
in addition to the one 
associated with $\HI$ reionization and moving the cutoff of the 1D flux power 
spectrum to lower $k$ modes.\footnote{In the simulations discussed in this work, we 
do not consider any high-$z$ $\HeII$ reionization model, but this could be 
easily added using the same formalism applied to $\HI$ reionization 
\citep[see][]{Onorbe:2017}.}
In any case, current observations of 
the 1D flux power spectrum at $z=5$ and $z=5.4$ disfavor
high-redshift $z\gtrsim9$ reionization models, far away from Planck
constraints on $\tau_e$, even for standard galaxy-driven heat injection.
We have shown that this is due to 
the dependence of the 1D flux power spectrum cutoff on the timing of 
reionization because the pressure-smoothing scale retains memory of the thermal 
history.

\subsection{Prospects for Measuring the Power Spectrum at \lowercase{$z\simeq 6$}}
\label{ssec:ps1dz6}

\begin{figure*}
\begin{center}
\includegraphics[angle=0,width=0.48\textwidth]{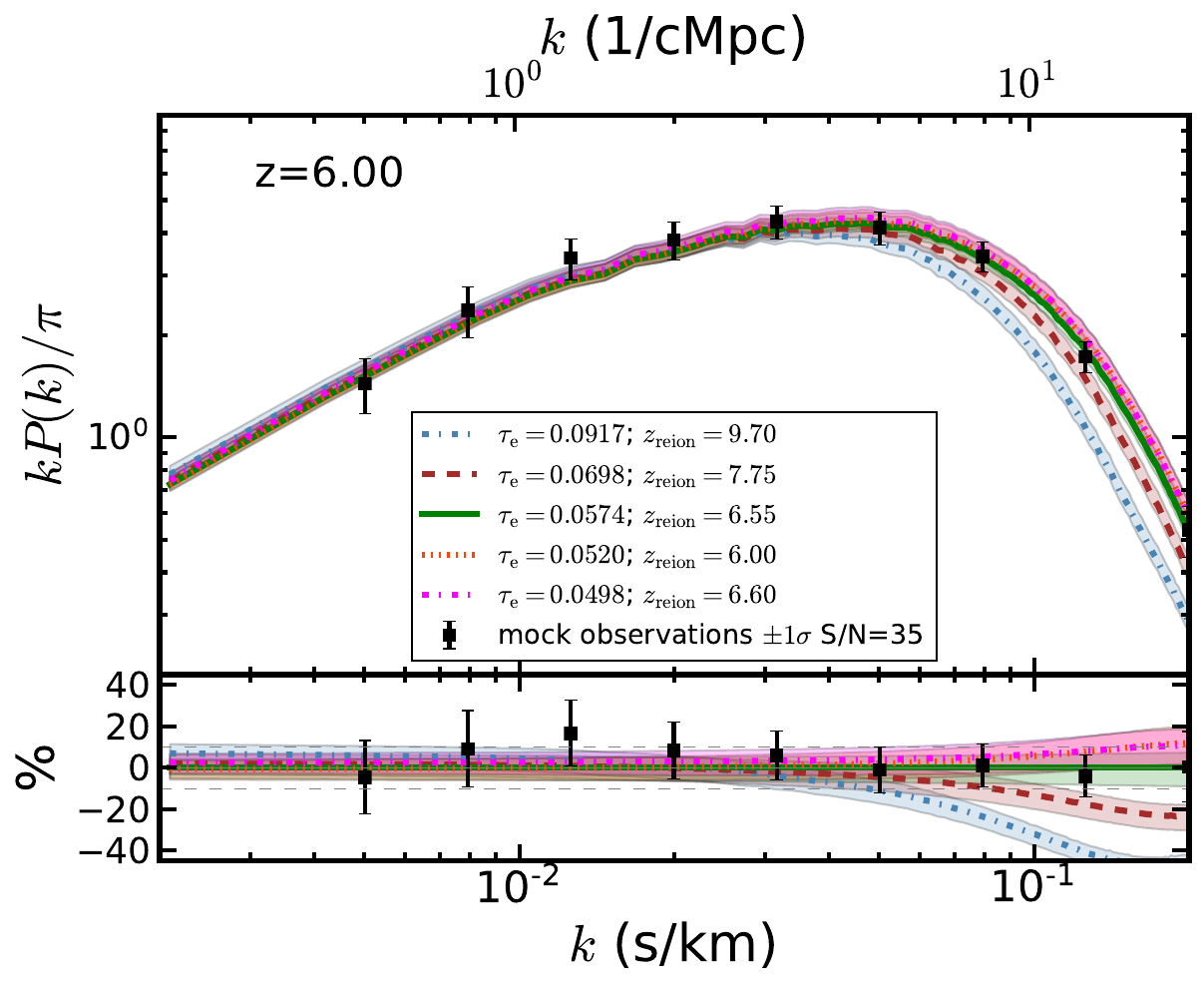}
\includegraphics[angle=0,width=0.48\textwidth]{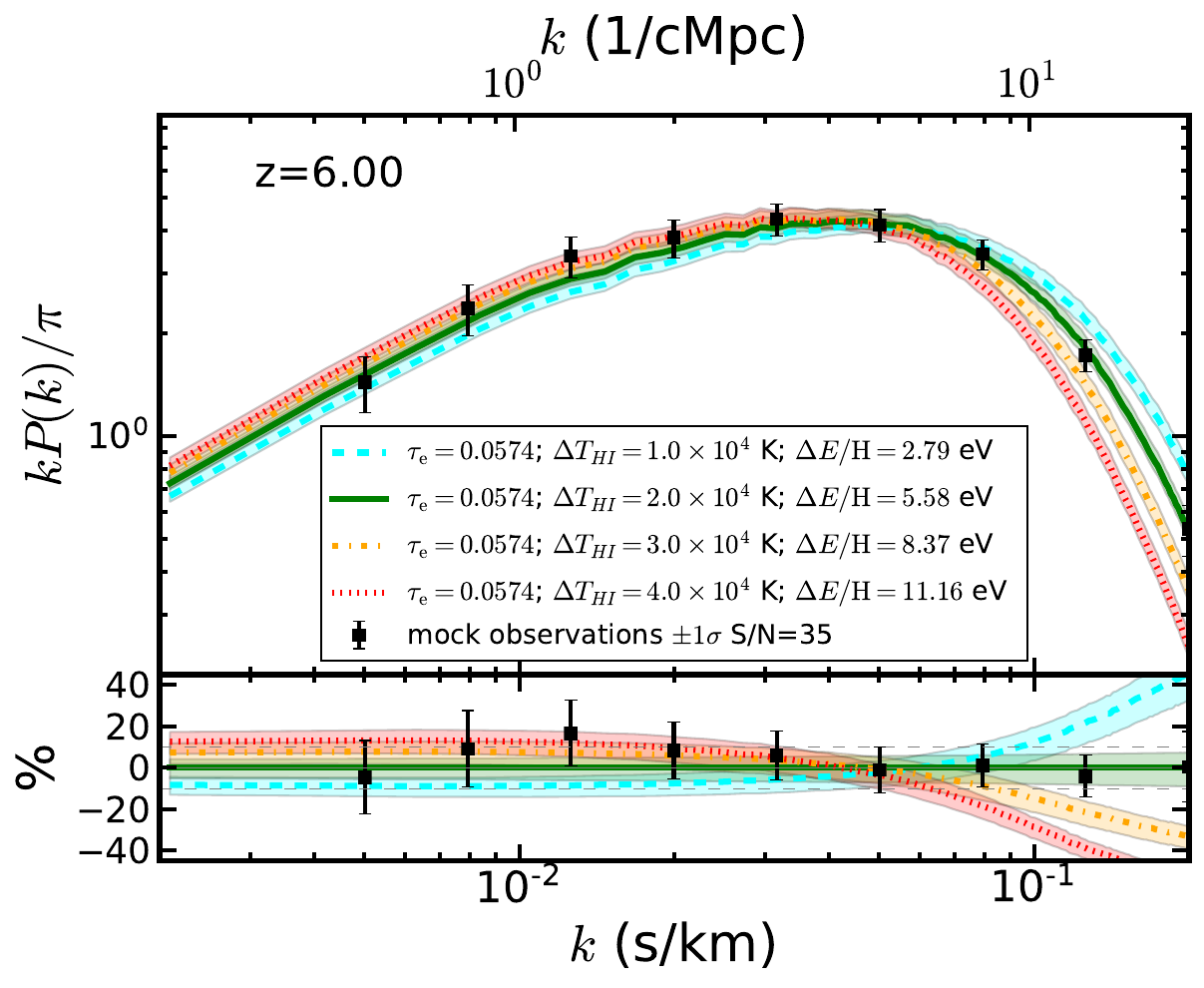}
\end{center}
\caption{Effect of reionization on the 1D flux power spectrum at $z=6$. Left 
panel: 1D flux power spectrum at $z=6$ for simulations that differ in their 
$\HI$ reionization history: EarlyR, MiddleR, LateR, MiddleR-fast, and VeryEarlyR, 
but share the same heat input. Right panel: 1D flux power spectrum at $z=6$ for 
simulations in which reionization happened at the same time but that differ in the heat 
input during $\HI$ reionization: Middle-cold, MiddleR, MiddleR-warm, and 
MiddleR-hot. Black squares stand for MiddleR-hot mock observations using 
high-resolution spectra of 10 quasars at $z\sim 6.1$ (using a path length 
equivalent of $\Delta z=0.5$ per quasar).
Color bands show the variation in the 1D flux power spectrum due to
one-sigma changes in the mean flux.
See text for more details.
\label{fig:ps1dz6}}
\end{figure*}

Motivated by the results in the previous subsection for the 1D flux power 
spectrum at $z=5.0,5.4$ and the increasing number of high-$z$ quasars that are 
uncovered by recent surveys 
\citep[e.g.][]{Banados:2014,Matsuaoka:2016,Banados:2016}, we also wish to 
explore the 1D flux power spectrum at  $z=6$,
and study the feasibility of making a power spectrum 
measurement at this redshift with current facilities. Whereas, as discussed 
previously, the power spectrum signal increases toward increasing redshift, the 
mean flux also begins to drop precipitously, lowering the signal-to-noise (${S\slash N}$) ratio level of the 
quasar spectra, thus increasing the importance of noise in the power spectrum 
measurement.
At $z=6.0$ we have 
assumed a mean flux value of $\mean{F}=0.011$ ($\mean{\tau_{\rm eff}}=4.5$), which is consistent
with the latest measurements
of the effective optical depth at this redshift
\citep{Becker:2015,DAloisio:2016}.
We have also assumed that the $1\sigma$
error on the mean flux at this redshift
will be $10\%$ of its value, which is a reasonable assumption given the large numbers of
$z\sim 6$ quasars recently discovered, $\sim 150$. 

To this end, we 
computed mock observations for the MiddleR simulation assuming 
high-resolution spectra (${S\slash N}$
of 35 per resolution element of $8$ km s$^{-1}$) 
of 10 
quasars at $z=6.3$ employing a path length equivalent of $\Delta z=0.5$ per 
quasar. 
For this we calculate the corresponding path length per quasar in cMpc 
at this redshift. We created random samples for the 10 quasars from the 
simulations and added noise realizations to each skewer. We then computed the mean 
power spectrum and subtracted off the average noise level. 
Results of one of these mock observations are shown as black squares in Figure~\ref{fig:ps1dz6}
along side with the error bars computed from a set of 50 mock observations.
At the $S\slash N$ considered, the measurement is dominated by cosmic variance.
The noise in the quasar spectra is 
very significant at this redshift, but 
the increase in overall power due to the decrease of the mean flux
still allows us to measure
of the power using a sample of 10 quasars.

The two panels of Figure~\ref{fig:ps1dz6} show the 1D flux power spectrum for 
the same simulations as discussed in Figure~\ref{fig:ps1dzreion} (left panel) and 
Figure~\ref{fig:ps1dheat} (right panel), but now at $z=6$. The color bands for 
each model indicate the variation in the 1D flux power spectrum due to $1\sigma$
changes in the mean flux. We can see that the overall scale of the flux power 
has increased compared to the values at lower redshift due to the decrease in 
mean flux. As discussed above, the power increases with redshift as the mean flux 
goes down because this amplifies the fluctuations. Note that the differences 
between the models have increased relative to $z\sim5$, especially for high-$k$
modes and for simulations where the heat injection is varied.
This is because at $z=6$, we are closer to 
reionization and therefore not only $\lj$, but also the other thermal 
parameters, $T_{0}$ and $\gamma$ (see Figure~\ref{fig:thermalhistory}),
are still affected by the details on how $\HI$ 
reionization happened \citep{McQuinn:2016}.

Data with a size and $S\slash N$ comparable to our assumed mock
are clearly within reach.
For example, about $5$ such quasar spectra already exist in public
telescope archives \citep{Becker:2015}, so a sample of $10$
would be accessible with modest
allocations of an $8$ m class telescope time.
We have shown that the differences between models are greater at $z=6$
than at lower redshift, compensating for the possibly lower $S\slash N$.
In order to understand how reionization heated the IGM 
and constrain both the reionization history and the heat
injected, we therefore need to push as far back into their
reionization epoch as possible,
where not only $\lj$, but also
$\gamma$ and $T_{0}$ still could have memory of reionization.

\section{Discussion}
\label{sec:disc}

\subsection{Convergence of the Results}
\label{ssec:converge}

While the resolution and method
employed in our \lyalpha{} simulations are 
currently state-of-the-art
for this type of analysis 
\citep{Lukic:2015,Onorbe:2017}, several aspects of our simulations call for 
caution. First, one should consider the level of convergence
of the 1D flux 
power spectrum at these high redshifts for our grid of simulations that have 
$L_{\rm box}=20$ Mpc h$^{-1}$
and $N=1024^3$. \citet{Lukic:2015} reported a careful 
convergence analysis of Nyx simulations of the \lyalpha{}
forest similar to those employed in this work, 
but they only explored $2\leq z \leq 4$. At $z=4$, the 1D flux power spectrum of 
their simulations were converged to $<5\%$ in terms of spatial resolution 
(missing power in large k modes) but $<8\%$ due to box size effects 
(overestimated power at large $k$ modes).
We show in the
Appendix~\ref{app:convergence} resolution and box size convergence results at 
$z=5$ and $z=6$ that reach similar conclusions, although approaching $\lesssim10\%$
convergence level at the $k\sim0.1$ s km$^{-1}$ modes, as typically used 
for power spectrum measurements,
and much better as we move to lower 
$k$ modes (larger scales). In any case, although convergence issues
have to be taken 
into account, they do not seem to change the general conclusions of this work. 
Similar results at these redshifts but for simulations using the Gadget code can 
be found in \citet{Bolton:2009} and \citet{Bolton:2017}.

Another relevant issue is that our simulations do not model 
galaxy formation.
Therefore we neglect any impact that the effect of 
stellar or black hole feedback could have on the IGM 
\citep[][]{Theuns:2002b,Kollmeier:2006,Desjacques:2006,TepperGarcia:2012}.
\citet{Viel:2013b} showed that at $z\sim3$ this effect could lead to differences 
of $\sim10\%$ in the 1D flux power spectrum. This is a very relevant issue that 
should be explored in more detail with current state-of-the-art feedback models. 
However, it is expected that the effects of feedback on the IGM should only decrease
at higher redshifts for two reasons: first, the \lyalpha{} forest is 
tracing lower gas densities at average locations of the universe 
\citep[see, e.g., Fig. 7 in][]{Lukic:2015},
and the smaller number 
of galaxies or black holes at these redshifts make it difficult for any feedback 
to alter the thermal state of the IGM.

Recent measurements of the \lyalpha{} optical depth at high redshift have found 
enhanced scatter at $z>5.5$ that exceeds what can be attributed to density 
fluctuations alone \citep{Fan:2006,Becker:2015}. It has been argued that they 
are driven by fluctuations in the radiation field \citep{Davies:2016,DAloisio:2016}, or the 
temperature field \citep{DAloisio:2015}, both of which may be inevitable 
byproducts of a patchy, extended, and late-ending reionization process. Still 
others have interpreted these fluctuations as evidence that reionization was 
actually driven by rare AGN \citep{Chardin:2015,Madau:2015,Khaire:2016}. 
In any case, the 
possible effect of UVB or temperature fluctuations is currently neglected in 
standard optically thin simulations like those we used in this work. 
Previous studies have shown that these effects are manifest on much larger scales, 
$\gtrsim10$ Mpc h$^{-1}$, that is,so much lower $k$ modes than those most sensitive to the thermal
state of the IGM 
\citep[e.g.][]{Abel:1999,Meiksin:2004,Cen:2009,Pontzen:2014,GontchoaGontcho:2014,Malloy:2015,DAloisio:2016}.
For example, \citet{DAloisio:2016} computed semi-numerical models with and without temperature 
fluctuations and showed that they did not generate small-scale power. 
Considering both temperature and UVB fluctuations is clearly
the direction in which the modeling needs to move forward. We plan to study
this in detail in the near future with self-consistent hydrodynamical simulations.

\subsection{Degeneracy with Cosmological Parameters and Warm Dark Matter}
\label{ssec:cosmo}

The 1D flux power spectrum depends not only on the thermal parameters of the IGM, 
but also on cosmological parameters. Here we have focused our analysis on one 
specific cosmological model, but we have checked that when one considers the 
range of parameters allowed by \citet{Planck:2015} the differences at the $k$ 
modes studied in this work are never larger than $8\%$
and the cutoff is unaffected 
\citep[see][for more details on these models]{Onorbe:2017}.

While changes of cosmological parameters within $1\sigma$ of the Planck constraints
result in minimal changes of the $z \sim 5-6$ power spectrum 
relative to the parameters governing reionization and its heating effect, 
this is not true when 
one considers dark matter particle properties, such as warm dark matter.  Their 
free-streaming horizon leads to a suppression of the small-scale power and 
therefore to a degenerate effect on the power spectrum with the thermal 
parameters \citep{Viel:2013a,Garzilli:2015}. This is 
equivalent to a 3D smoothing at high redshift that continues to decrease as 
nonlinearities increase. Therefore the small-scale
cutoff in warm dark matter models
moves from very low k modes (large scales) to higher k modes (smaller scale) 
at progressively lower redshifts.
This happens until the IGM becomes hotter and the IGM temperature (i.e. the $T_{0}$ and $\gamma$)
determines the position of the cutoff.
In order to provide reliable 
WDM constraints, it is therefore essential to marginalize out reionization nuisance 
parameters.
\citet{Garzilli:2015} highlighted the degeneracy between different 
warm dark matter masses and the temperature of the IGM. However, these authors did 
not discuss the extra degeneracy
resulting from the unknown redshift of reionization and its associated heat injection,
as we demonstrate in this work.
This could be relevant given that the fiducial value 
in their models is $z_{\rm reion}=12$ and the lowest reionization redshift 
considered in their grid, by including one simulation, was $z_{\rm reion}=8$. 
From their Bayesian analysis, \citet{Viel:2013b} and \citet{Garzilli:2015} 
obtained very low temperature values at $5<z<5.4$ (see 
Figure~\ref{fig:thermalhistory}). Although this is not in complete disagreement 
with our qualitative comparison of the measurements to our models, 
we caution that their simulations use a total of 
$2\times 512^3$ dark matter and gas particles within a periodic box of $L_{\rm 
box}=20$ cMpc h$^{-1}$\footnote{Note that \citet{Viel:2013b} use 4 N=$2\times 768^3$ and $L_{\rm 
box}=20$ cMpc h$^{-1}$ simulations to correct for resolution convergence.}.
Based on the resolution convergence tests presented in the
\citet[][see their Figure A4]{Bolton:2017} 
spectrum using the same code, these simulations could underestimate the power 
with $\sim 20\%$ error at the k modes most relevant to study the cut off of the 
power spectrum ($0.05 \lesssim$ k $\lesssim 0.1$ s/km). This could produce an 
artificial increase of the cutoff just due to resolution.

\subsection{Comparison to Previous Work}
\label{ssec:comparison}

\citet{Nasir:2016} used hydrodynamical simulations with a total of $2\times 
512^3$ dark matter and gas particles within a periodic box of $L_{\rm box}=10$ 
cMpc h$^{-1}$ to discuss the possibility of constraining the thermal history of the IGM 
during $\HI$ reionization by studying the 1D flux power spectrum at $z=5$ from 
cosmological hydrodynamical simulations.
To simulate different reionization histories, they adopted 
an approach different from ours. Namely they applied a multiplying 
factor to the \citet{Haardt:2001} photoheating rates ($A\times \dot{q}\times 
\Delta^{B}$, where $\Delta$ is the specific overdensity of that cell). They also used 
simple cutoffs of the \citet{Haardt:2001} UVB rates at various redshifts to model
different reionization timing.
Using these simulations, they showed the effect on the 1D flux power spectrum of 
different thermal histories in which they only changed the timing of $\HI$ 
reionization
and tried to study its degeneracy with the temperature of the IGM, $T_{0}$. 
Interestingly, these authors found that this degeneracy between the timing of 
reionization and the temperature can in fact be broken using different scales of 
the 1D flux power spectrum. They find that the $z=5$ 1D flux power spectrum is 
more sensitive to the timing at $0.03< k <0.13$ s km$^{-1}$ scales. Our analysis also 
shows that these scales are the most sensitive to the timing of reionization 
(see Figures~\ref{fig:ps1dzreion}), 
but it also indicates that information about the lower $k$ modes
will be crucial in 
order to break the degeneracy between the timing of reionization and other 
parameters, the temperature at mean density, $T_{0}$ or the mean flux at that 
specific redshift.

\citet{Nasir:2016} attempt to quantify the effect of a different reionization 
timing using a new parameter, $u_{0}$, which is an integral in time of the 
heating rate per proton mass of the simulation at mean density.\footnote{It is 
defined as $u_{0}(z)=\int_{z}^{z_{\rm reion}} \frac{\sum_{i} 
n_{i}\dot{q}_{i}}{\bar{\rho}} \frac{dz}{H(z)(1+z)}$, where $\bar{\rho}$ is the 
mean background baryon density and $n_{i}$ and $\dot{q}$ stand for the density 
and photoheating rates of the following species: $i= [\HI,\HeI,\HeII]$.}
With this parameter, the authors try to quantify the pressure-smoothing
scale in their models.
In order 
to facilitate comparison with their work we have computed the value of $u_{0}$ at 
$z=4.9$ for our simulations and included them in Table~\ref{tab:sims}. However, 
we caution about using this parametrization as $u_0$ measures the heating just at 
mean density.
Therefore this parametrization will be valid as long as the 
the reionization models do not have density-dependent heating.
Models with heating rates that depend on the density 
but normalized at mean density \citep[e.g.][]{Becker:2011,Becker:2013} share 
the same $u_0$ value, but have a different power spectrum at the $k$ modes
that are
more sensitive to the thermal history, $k>0.03$ s km$^{-1}$, as they in fact
have different pressure-smoothing scales.

\section{Conclusions}
\label{sec:conc}

In this paper we have used state-of-the-art hydrodynamical simulations that 
allowed us to self-consistently model different reionization models.
We present an ensemble of simulations consistent with the latest 
measurements of the Thompson-scattering optical depth, $\taue$ recently reported 
by Planck \citep{Planck:2016a,Planck:2016b}. These models are defined by when 
reionization happened, $z_{\rm reion}$, and how much heat was injected into the 
IGM during reionization, $\Delta T$. Our simulations show that although 
by $z\sim6$ the 
temperature of IGM gas has mostly forgotten about reionization heat injection, the pressure-smoothing scale
at these redshifts 
depends sensitively on how and when
reionization occurred. This is because both the 
cooling and dynamical times in the rarefied IGM are long, comparable to the 
Hubble time, and therefore memory of $\HI$ reionization is retained
\citep{Rorai:2013,Kulkarni:2015,Onorbe:2017}. 
We have found a degeneracy in the 
pressure-smoothing scale at $z<6$ between when reionization occurred and the 
amount of heat injected during reionization. For a fixed reionization history,
the pressure-smoothing scale increases as we increase the heat injection. 
Similarly, the pressure-smoothing scale increases with the redshift
of reionization in models with a fixed amount of heat injection.

In order to investigate the effects of these different thermal histories on the 
properties of the \lyalpha{} forest, we compute the \lyalpha{} 1D flux 
power spectrum at $z\sim 5-6$ for our simulation ensemble.
Pressure smoothing 
damps out small-scale fluctuations in the IGM, while thermal vibrations of IGM 
gas Doppler broadens \lyalpha{} forest lines, further reducing the amount of 
small-scale structure. Both of these effects combine to produce a prominent 
small-scale (high-k) cutoff in the \lyalpha{} 1D flux power spectrum 
\citep{Zaldarriaga:2001,Peeples:2010a}. We have found that at these high redshifts, the 
1D flux power spectrum is especially sensitive to the pressure-smoothing 
scale of the IGM and not only its temperature. Therefore extant thermal 
signatures from reionization can be detected by analyzing the \lyalpha{}
forest power spectrum at these redshifts.

We have also conducted a first qualitative comparison of the
1D flux power spectrum measurements at $z=5-5.4$ made by 
\citep{Viel:2013a} with our simulation ensemble.
Taking Planck 
constraints on reionization at face value, we have shown that
models with a fiducial heat input during $\HI$ reionization consistent with 
standard galaxy-driven reionization 
models are sufficient to explain the observations. We work on a more complete analysis
of this in the near future, with a larger simulation grid,
marginalizing out all the different relevant parameters, including the mean 
flux, and improving upon the reionization modeling.

We have also presented a feasibility study of performing a similar measurement at $z=6$ 
creating mock observations that assumed a realistic sample of quasars at this 
redshift both in terms of sample size and $S\slash N$.
We found that combining 10 quasars should be enough to distinguish 
between different thermal histories of the IGM. Our results indicate that
quasar spectra at high redshift can not only be useful to 
constrain when reionization happened 
via the study of \lyalpha{} opacity measurements \citep[e.g.][]{Fan:2006,Becker:2015},
but also to understand the thermal history of the universe.
This is the small-scale structure measured from high-resolution
spectra can be used to understand the thermal history of the universe, further constraining
the timing and heat injection by reionization. 
Taking into account that 
there are only a few direct observational probes of reionization currently
available, we think that 
it is important to push in this direction in the near future.

In this regard, pushing these measurements at this and higher redshifts ($z\sim6$) will be 
crucial to improve the power of the \lyalpha{} forest to constrain $\HI$ 
reionization. The new recent fivefold increase in the number of bright quasars 
at $z>5$ is obtained from deep wide-field optical/IR surveys like CFHQS \citep{Willott:2010}, 
dark energy survey \citep[DES,][]{Reed:2015},
ESO public surveys \citep[VST/KiDS and VISTA/VIKING][]{Venemans:2015b},
and Pan-STARRS1 \citep{Banados:2014,Venemans:2015a}. Currently, the total number
$z > 5.5$ quasars available for study is $\sim173$, so if all could be used
to study the cutoff of the \lyalpha{} forest power spectrum would reduce
the errors by a factor of $\sim 4$ compared with current measurements.
Moreover, large datasets of medium- and high-resolution \lyalpha{} spectra are already available
at lower redshifts \citep[$z\lesssim4$, e.g.][]{OMeara:2015,Irsic:2017} that combined with the large redshift
data can provide a more comprehensive constraint of reionization.
Therefore this requires 
starting to focus now on improving the theory to exploit this increased precision. 
While modeling the 1D flux power spectrum with hydrodynamical simulations at sufficiently high accuracy 
is an incredible computational challenge, 
the advent of high-performance computing power and the high scalability of Nyx
has allowed us to significantly improve the accuracy of our predictions
in recent years.

Finally, given that there are
few observables that are sensitive to the thermal state of baryons at the earliest redshifts,
the 1D flux power spectrum at $z \gtrsim 5$ offers a unique opportunity to explore not only $\HI$ reionization, but
also constrain other physical scenarios that alter the thermal history of the IGM,
for example, models that alter the thermal state of the IGM 
via X-ray pre-heating coming from 
starburst galaxies, supernova remnants or miniquasars 
\citep{Oh:2001,Glover:2003,Madau:2004,Furlanetto:2006,Madau:2017}, dark matter annihilation 
or decay \citep[][]{Liu:2016}, cosmic rays \citep{Samui:2005}, blazar heating \citep{Chang:2012,Puchwein:2012},
broadband intergalactic dust absorption \citep{Inoue:2008}
or high-$z$ exotic reionization scenarios driven by Population 
III stars \citep[e.g.][]{Manrique:2015}.

\acknowledgments
We thank M.~White, J.~Miralda-Escud\'e and the members of the ENIGMA group at 
the Max Planck Institute for Astronomy (MPIA) for helpful discussions.

ZL was partially supported by the 
Scientific Discovery through Advanced Computing (SciDAC) program funded by U.S. 
Department of Energy Office of Advanced Scientific Computing Research and the 
Office of High Energy Physics. Calculations presented in this paper used the 
hydra cluster of the Max Planck Computing and Data Facility (MPCDF, formerly 
known as RZG) MPCDF is a competence center of the Max Planck Society located in 
Garching (Germany). We also used resources of the National Energy Research 
Scientific Computing Center (NERSC), which is supported by the Office of Science 
of the U.S. Department of Energy under Contract No. DE-AC02-05CH11231. 
This work made 
extensive use of the NASA Astrophysics Data System and of the astro-ph preprint 
archive at arXiv.org. This work is based on observations obtained with Planck 
(http://www.esa.int/Planck), an ESA science mission with instruments and 
contributions directly funded by ESA Member States, NASA, and Canada.


\appendix

\section{Numerical Convergence}
\label{app:convergence}

\begin{figure*}
 \begin{center}
 \includegraphics[angle=0,width=0.45\textwidth]{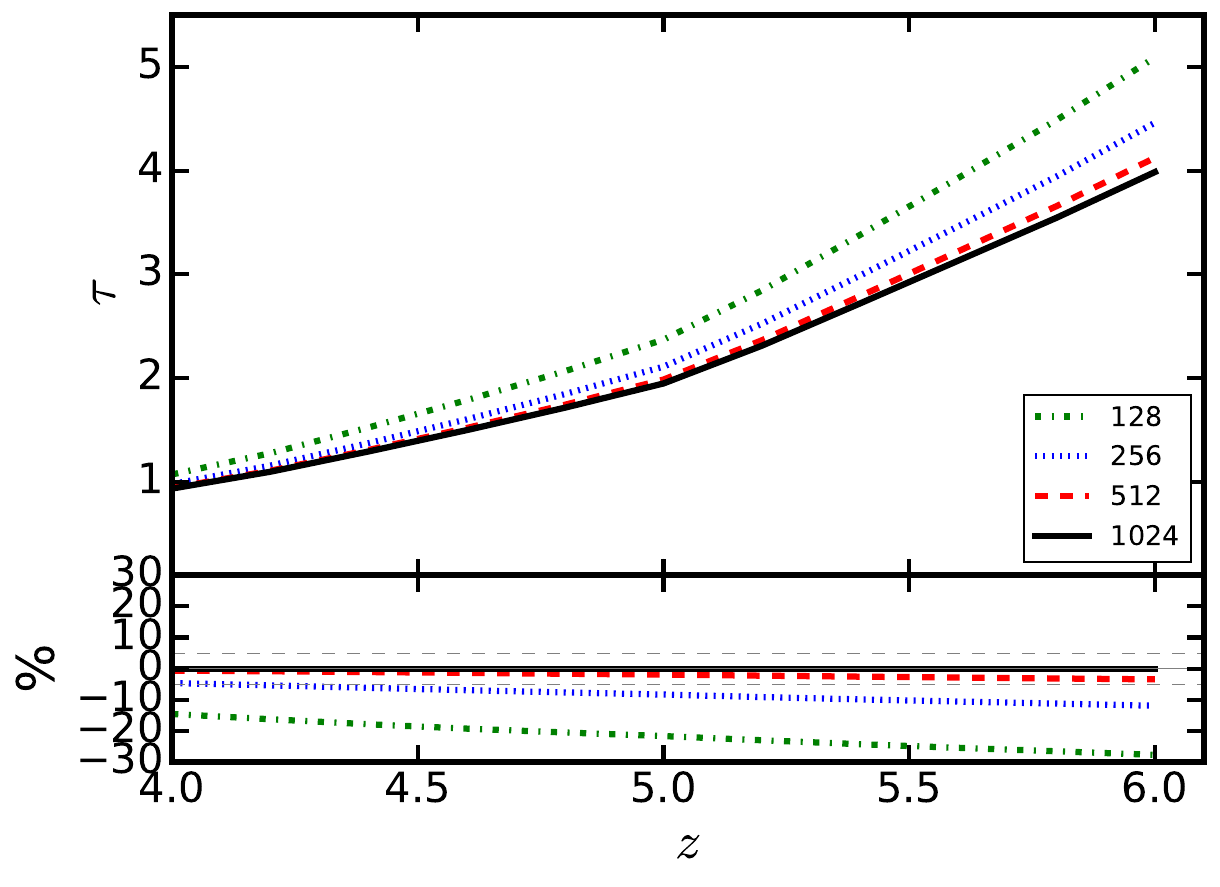}
 \includegraphics[angle=0,width=0.45\textwidth]{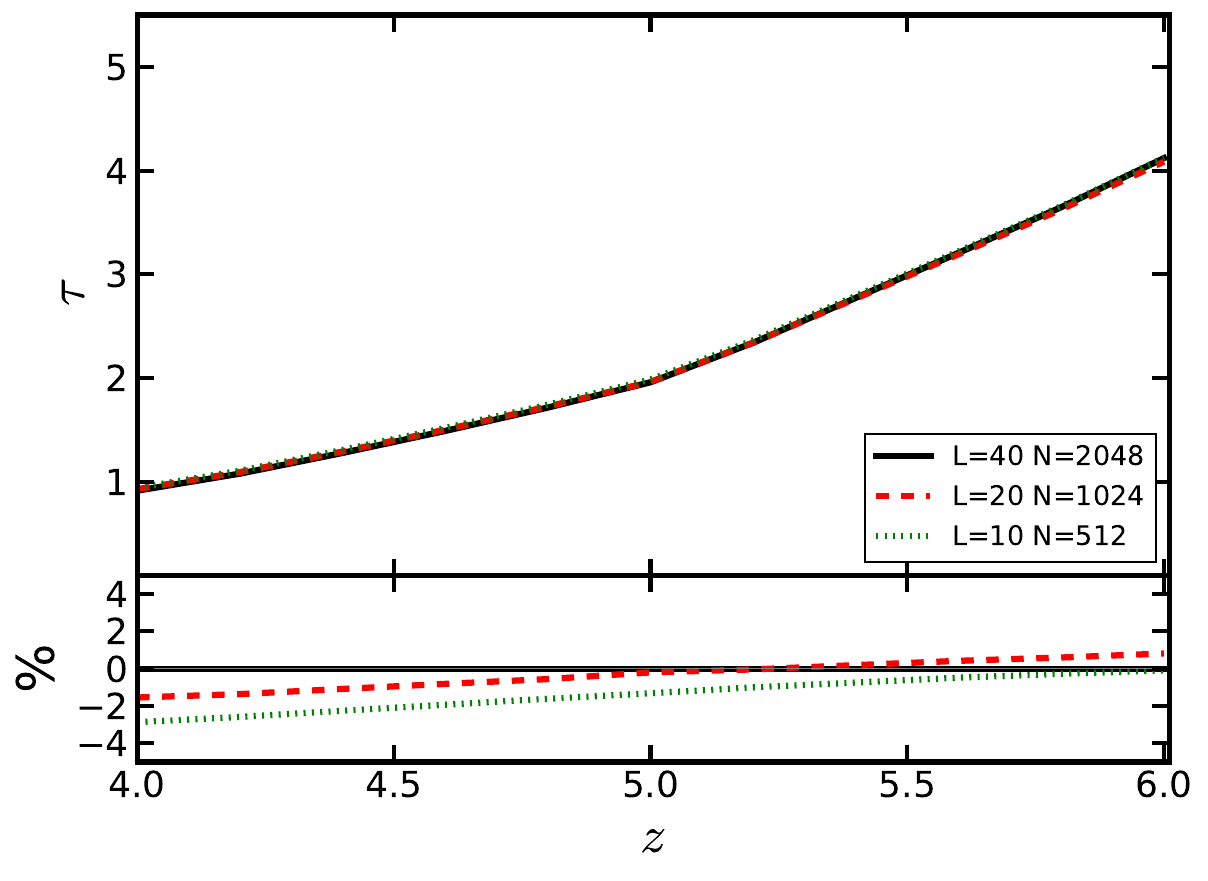}
  \end{center}
 \caption{Mean optical depth ($\mean{\tau_{\rm eff}}$) convergence results from $z=4.0$ up to $z=6.0$. Left panel: simulations with 
 a fixed box size ($L_{\rm box}=10$ Mpc h$^{-1}$) and different spatial resolution, $\Delta x=78$ (dot-dashed green line), $39$ (dotted blue line), $20$ (dashed red line), and 
$10$ kpc h$^{-1}$ (black line). Right panel: simulations with a fixed spatial resolution 
($\Delta x\sim 20$ kpc h$^{-1}$) and different box size, $L_{\rm box}=10$ (dotted green line), $20$ (dashed red), $40$ Mpc h$^{-1}$ (black line).
In both panels the red-dashed lines correspond to the simulations discussed in this paper.
 \label{fig:tauconvergence}}
 \end{figure*}

 \begin{figure*}
 \begin{center}
 \includegraphics[angle=0,width=0.45\textwidth]{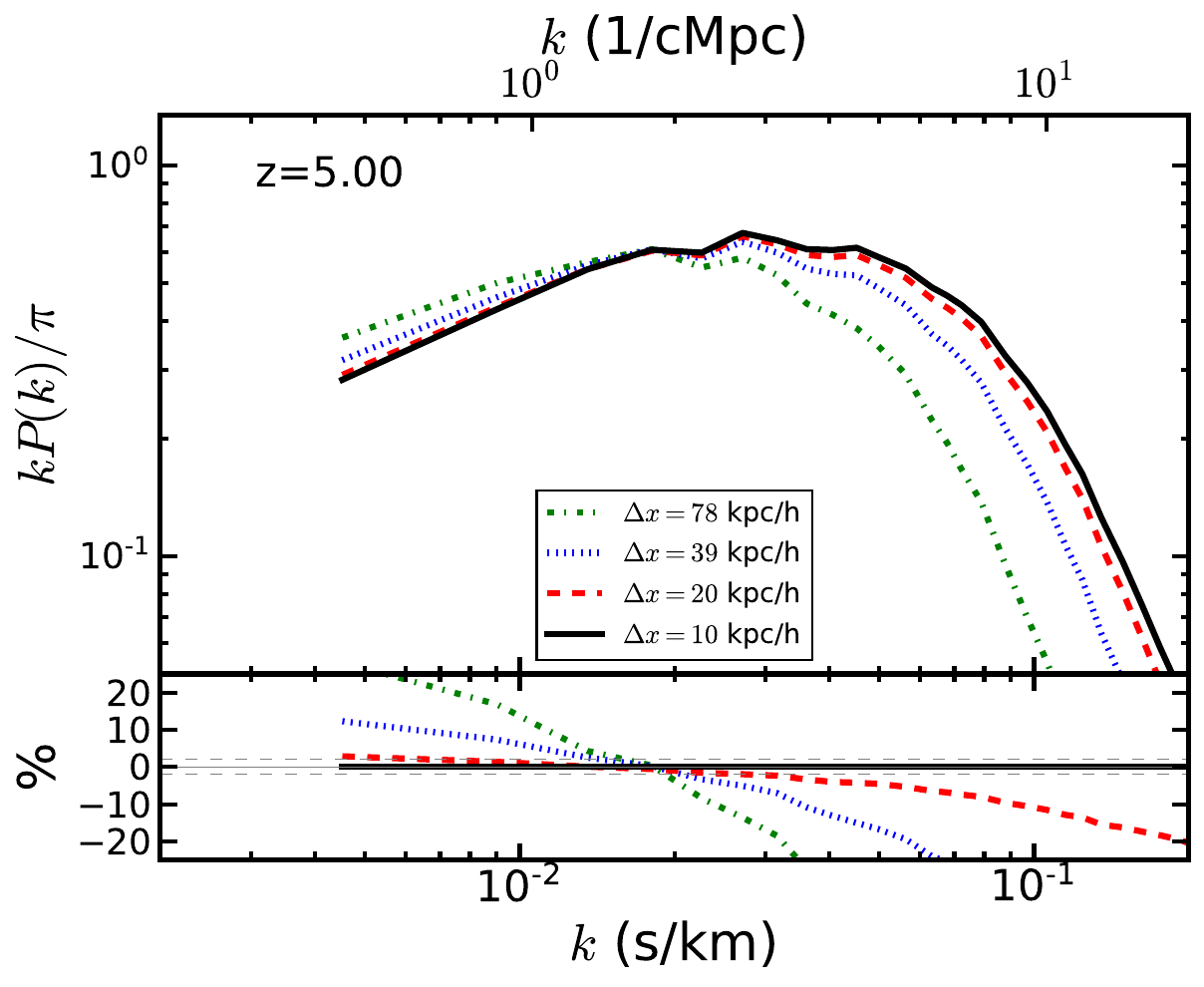}
 \includegraphics[angle=0,width=0.45\textwidth]{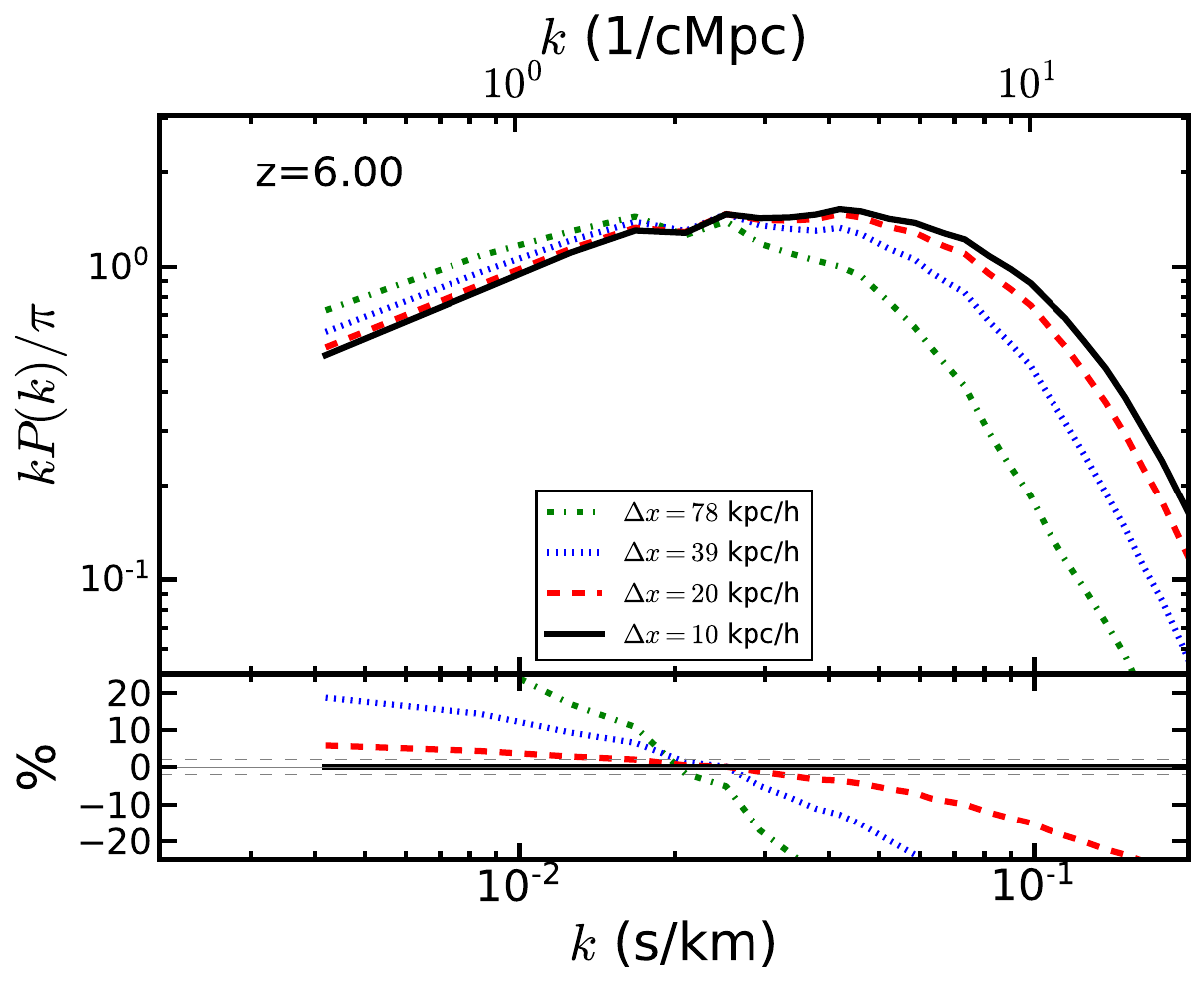}\\
 \includegraphics[angle=0,width=0.45\textwidth]{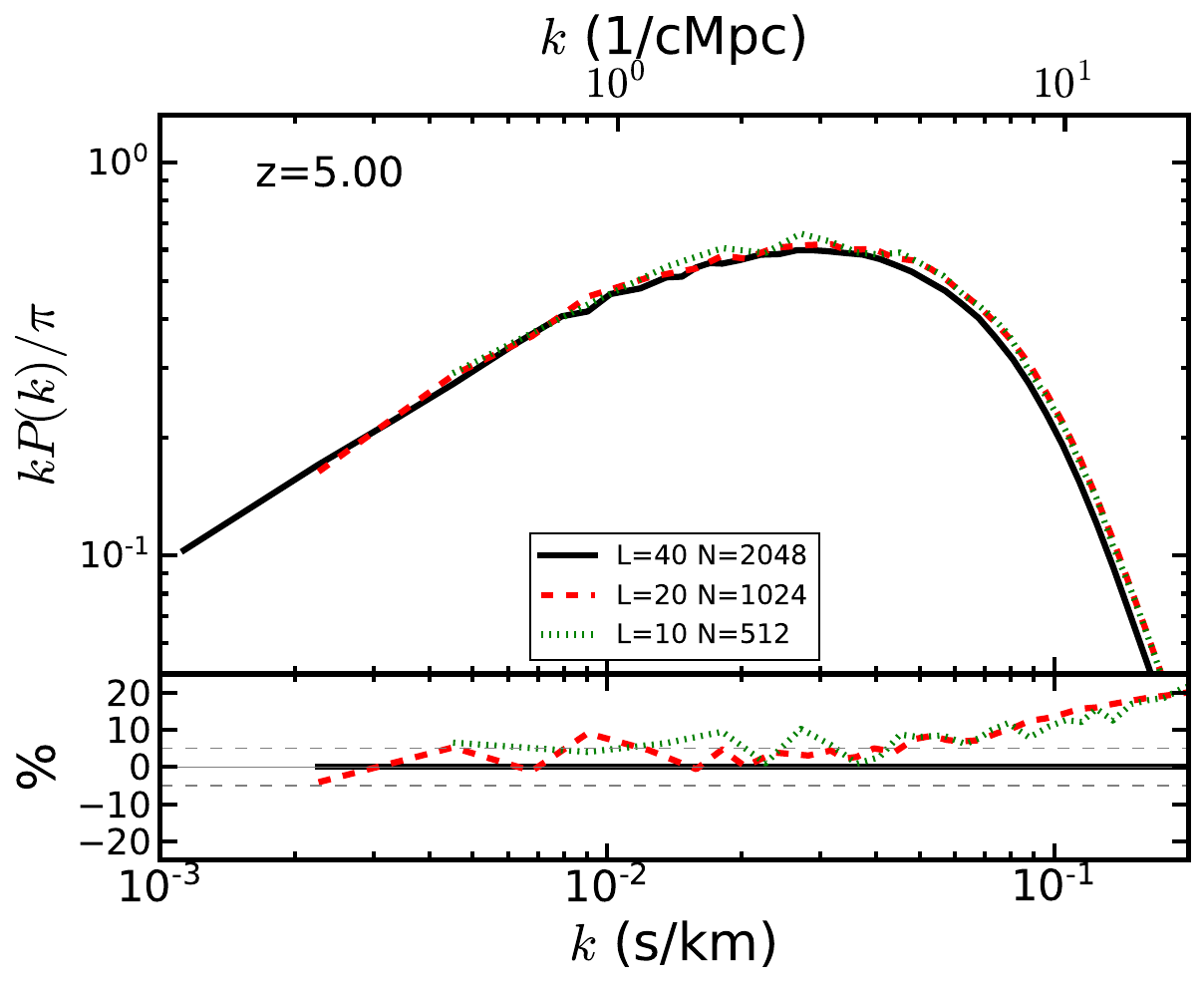}
 \includegraphics[angle=0,width=0.45\textwidth]{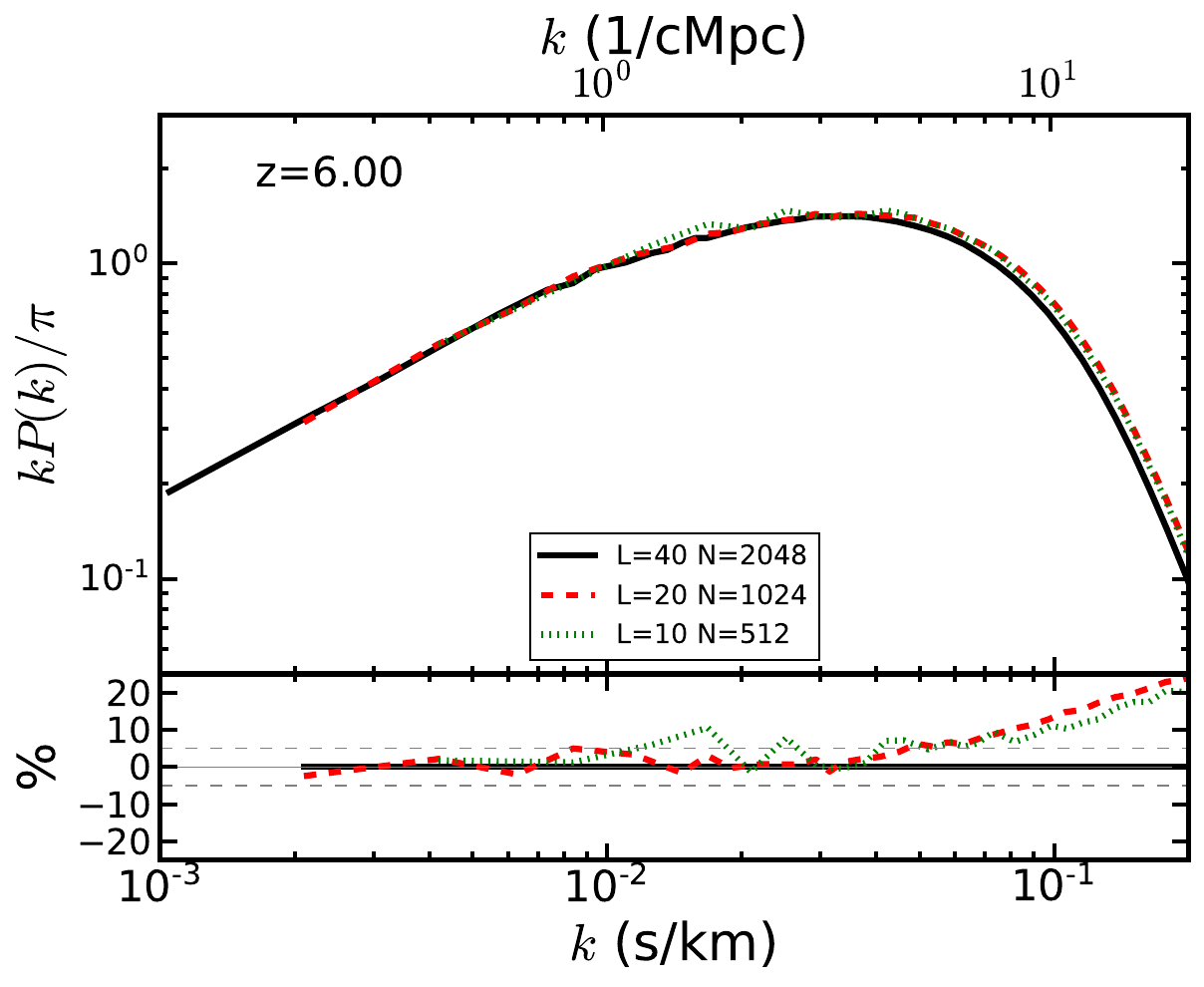}
  \end{center}
 \caption{Convergence results for the 1D flux power spectrum at $z=5$ (left column) and the right column at $z=6$ (right column). 
 The upper panels present results for simulations with a fixed box size ($L_{\rm box}=10$ Mpc h$^{-1}$)
 and different spatial resolution, $\Delta x=78$ (dot-dashed green line), $39$ (dotted blue line), $20$ (dashed red line), and 
$10$ kpc h$^{-1}$ (black line). The lower panels show the 1D flux power spectrum for simulations with a fixed spatial resolution 
($\Delta x\sim 20$ kpc h$^{-1}$) and different box size, $L_{\rm box}=10$ (dotted green line), $20$ (dashed red), $40$ Mpc h$^{-1}$ (black line).
In all panels the red-dashed lines correspond to the simulations discussed in this paper.
 \label{fig:ps1dconvergence}}
 \end{figure*}

In this section we discuss the convergence 
tests with spatial resolution and box size for the two fundamental quantities
studied in this paper, the mean optical depth and the 1D flux power spectrum.
To study the effects of spatial resolution, we have run 4 simulations with the same thermal
history (e.g., same UVB; MiddleR) and box size of $L_{\rm box}=10$ Mpc h$^{-1}$ but increasing numbers of
resolution elements: $128^3$ 
(dot-dashed green), $256^3$ (dotted blue),
$512^3$ (dashed red), and $1024^3$ (black). 
To simplify the comparison, simulations performed in the same box size
share the same large-scale modes, the only difference being that
higher resolution runs have more modes sampled on small scales.
These simulations have a cell size of $78$, $39$, $20$, and 
$10$ kpc h$^{-1}$, respectively, and therefore the $512^3$ run has the same spatial resolution
as the simulations discussed in this work ($L_{\rm box}=20$ Mpc h$^{-1}$ and $1024^3$ cells).
We also ran one more simulation with the same thermal history 
(MiddleR), a box size of $L_{\rm box}=40$ Mpc h$^{-1}$ and $2048^3$ cells
in order to study box size effects. We compare this simulation with 
two other runs with the same spatial resolution, but decreasing box size:
the $L_{\rm box}=20$ Mpc h$^{-1}$ - $N_{\rm cell}=1024^3$ run, which correspond to the 
simulations used in this work,
and the $L_{\rm box}=10$ Mpc h$^{-1}$ - $N_{\rm cell}=512^3$ run also used in the spatial
resolution study.

The evolution of the mean optical depth $\mean{\tau_{\rm eff}}$ for all these simulations
is shown in Figure~\ref{fig:tauconvergence}.
We computed its evolution directly from the 
simulation mean flux, $\mean{\tau_{\rm eff}}=-\ln \mean{F}$,
without any rescaling of the photoionization rate.
Thus, all simulations use exactly the same photoionization rates at all redshifts.
The left panel shows
the convergence of the mean optical depth as we increase the spatial resolution,
while the right panel shows the convergence for different box sizes.
The simulations discussed in this work (red dashed lines in both panels)
show a convergence level below  $<5\%$ 
between $4\leqslant z\leqslant6$
both in terms of spatial resolution and box size.

Figure~\ref{fig:ps1dconvergence} shows the convergence tests of 
the 1D flux power spectrum at redshift $z=5$ (left column) and $z=6$ (right
column) for the same simulations. 
For this test, we rescaled the mean flux of all the simulations
to the same value. We used the fit between mean flux
and redshift suggested
in \citet{Onorbe:2017}, 
obtained using a wide range of data sets between $0<z<6$,
but the exact values employed do not change our conclusions.
For the resolution tests (upper row) and the box size tests
(lower row), we find a $\lesssim10\%$ level of
convergence for $k$ modes lower than $\sim0.04$ s km$^{-1}$ 
in the simulations with the same resolution and box size used
in this work (red dashed lines).
The error in these modes is mainly
driven by box size effects as the resolution tests show
a better convergence. However, for modes more relevant to
study the thermal cutoff ($0.04< k<0.1$ s km$^{-1}$)
we find a $10\%$ convergence level
at $z=5$ and $15\%$ at $z=6$ that is mainly driven by spatial
resolution effects.
Note that this quoted convergence
level is the worst level at the highest $k$ mode, 
$0.1$ s km$^{-1}$, but it decreases as we move to lower $k$ values. 
It is also very relevant to indicate that spatial resolution
effects can only move the cutoff of the 1D flux power spectrum
to higher $k$ values as we increase the resolution. 
Similar results at these redshifts but for simulations
using the Gadget code
can be found in \citet{Bolton:2009b} and \citet{Bolton:2017}.
Convergence results at lower redshifts
for the same code used in this paper along
with a more detailed discussion can be found
in \citet{Lukic:2015}.


\bibliography{apj-jour,obsplanck}    

\label{lastpage}

\end{document}